\documentclass[%
 preprint,
 amsmath,amssymb,
 aps,
]{revtex4-2}

\usepackage{graphicx}% Include Figure files
\usepackage{dcolumn}% Align table columns on decimal point
\usepackage{bm}% bold math
\begin{document}

\preprint{APS/123-QED}

\title{Spatiotemporal Control of Two-Color Terahertz Generation}% Force line breaks with \\

\author{T. T. Simpson}
 \email{tsim@lle.rochester.edu}
\author{J. J. Pigeon}
\author{M. V. Ambat}
\author{K. G. Miller}
\author{D. Ramsey}
\author{K. Weichman}
\author{D. H. Froula}
\author{J. P. Palastro}
 \email{jpal@lle.rochester.edu}
\affiliation{%
 University of Rochester, Laboratory for Laser Energetics,  Rochester, New York 14623, USA
}%

\date{\today}% It is always \today, today,

\begin{abstract}
A laser pulse composed of a fundamental and properly phased second harmonic exhibits an asymmetric electric field that can drive a time-dependent current of photoionized electrons. The current produces an ultrashort burst of terahertz (THz) radiation. When driven by a conventional laser pulse, the THz radiation is emitted into a cone with an angle determined by the dispersion of the medium. Here we demonstrate that the programmable-velocity intensity peak of a spatiotemporally structured, two-color laser pulse can be used to control the emission angle, focal spot, and spectrum of the THz radiation. Of particular interest for applications, a structured pulse with a subluminal intensity peak can drive highly focusable, on-axis THz radiation.

%The conical emission can complicate collection and focusing of the radiation for use in applications.

\end{abstract}

\maketitle

Terahertz (THz) sources impact a range of scientific disciplines as probes for elucidating ultrafast dynamics and as pumps for accessing new states of matter. Residing in the band between microwave and optical frequencies, single-cycle THz sources satisfy the seemingly contradictory requirements of being both ultrafast ($\sim$1 ps) and quasi-DC. As probes, these sources have provided insights into carrier dynamics in semiconductors \cite{ulbricht2011carrier}, ferromagnetic phase transitions \cite{li2022ultrafast}, the conductivity of warm-dense matter \cite{chen2021ultrafast}, and the internal motion of biological molecules \cite{markelz2022perspective}. As pumps, THz pulses can induce exotic phase transitions \cite{disa2023photo} or exert strong magnetic or relativistic forces at intensities that are orders of magnitude lower than near-infrared laser pulses \cite{fulop2020laser}. Each of these applications has motivated continued development of high-power, well-characterized, and stable sources \cite{lewis2014review}. Plasma-based sources, in particular, benefit from a nearly limitless damage threshold, enabling the use of high-intensity laser pulses as drivers \cite{LeemansCTR2003, kim2007terahertz,antonsen2007excitation,Albert_2016,hafez2016intense,miao2017high,LiaoRelLaserFoil2020}.

\begin{figure}
\includegraphics[scale = 0.95]{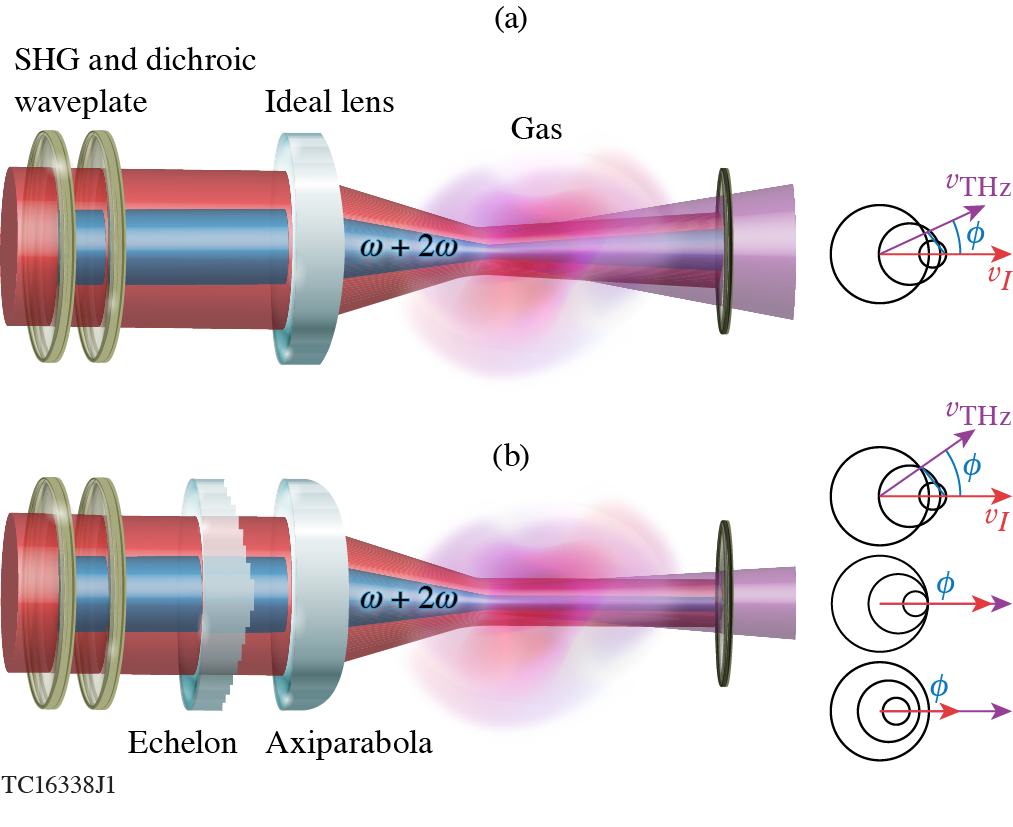}
\caption{Schematics of two-color THz generation schemes. (a) A conventional design. A two-color laser pulse is focused by an ideal lens, providing little to no control over the ionization front velocity or THz emission angle. (b) An ultrafast flying focus design. A two-color laser pulse is focused by an axiparabola, which extends the focal range, and passes through an echelon, which imparts a radial group delay designed to produce a desired ionization front velocity and THz emission angle. The flying focus optical assembly can create a focus (and corresponding ionization front) that move slower than (bottom), equal to (middle), or greater than (top) the THz phase velocity. For illustrative purposes the ideal lens, echelon, and axiparabola are shown in transmission, but these optics would typically be used in reflection.}
\label{fig:f1}
\end{figure}

Of the plasma-based sources, the ``two-color" technique routinely delivers well-characterized, single-cycle THz pulses at high-repetition rates \cite{dai2009coherent,babushkin2010ultrafast,kim2012high,clerici2013wavelength,de2015boosting,zhang2016controllable,yoo2016generation,andreeva2016ultrabroad,zhang2018manipulation,yoo2019highly,koulouklidis2020observation,buldt2021gas}. In two-color THz generation, a laser pulse composed of a fundamental and second harmonic ionizes a low-density gas \cite{kim2007terahertz,kim2008coherent} [Fig. \ref{fig:f1}(a)]. When the two harmonics are appropriately phased, the ionization creates a transient electron current that drives an ultrashort burst of THz radiation. The radiation constructively interferes in a Cherenkov-like cone with an angle $\phi$ determined by the velocity of the ionization front $v_I$ and the phase velocity of the THz $v_T$ \cite{johnson2013thz}:
\begin{equation}\label{eq:phi}
    \phi = \arccos(v_T/v_I).
\end{equation}
The velocity of the ionization front depends on the phase and group velocities of the two harmonics, which are linearly constrained by dispersion and evolve nonlinearly during propagation. This makes the emission angle $\phi$ difficult to control, especially when the gas species and density are used to optimize other properties of the radiation. Moreover, the conical emission complicates collection and focusing of the radiation for use in applications. 

Spatiotemporal pulse shaping can be used to control nonlinear propagation and the velocity of an ionization front independent of material dispersion \cite{palastro2018ionization, palastro2020dephasingless,simpson2020nonlinear,franke2021optical}. As an example, ``flying focus" techniques control the time and location at which different temporal slices, annuli, or frequencies within a laser pulse come to focus to produce a programmable-velocity intensity peak that travels many Rayleigh ranges with a near-constant profile \cite{froula2018spatiotemporal,sainte2017controlling,palastro2020dephasingless,simpson2020nonlinear,simpson2022spatiotemporal}. With sufficient intensity, the peak can trigger an ionization front that travels at a velocity $v_I \approx v_f$, where $v_f$ is the velocity of the moving focus \cite{palastro2018ionization,turnbull2018ionization,simpson2020nonlinear}. These arbitrary-velocity ionization fronts have been demonstrated experimentally \cite{turnbull2018ionization,franke2019measurement} and proposed for applications including extreme ultraviolet generation through photon acceleration \cite{howard2019photon,franke2021optical} and intense optical pulse generation using Raman amplification \cite{turnbull2018raman}.

In this Letter, we show that programmable-velocity ionization fronts driven by space-time structured, two-color laser pulses can be used to control the emission angle, focal spot, and spectrum of THz radiation. The velocity of the ionization front is tuned using a recently demonstrated optical configuration that creates ultrashort flying focus pulses \cite{palastro2018ionization,pigeon2023}. Adjusting the velocity of the ionization front provides access to three distinct regimes of THz generation. Matching the velocity of the ionization front to the THz phase velocity, $v_I = v_T$, produces a focusable, single-cycle THz pulse that constructively interferes on-axis ($\phi = 0$). Setting the ionization front velocity slightly lower than the THz phase velocity, $v_I \lesssim v_T$, also produces a focusable, on-axis THz pulse but with a longer duration and narrower spectrum. Finally, an ionization front velocity greater than the THz phase velocity, $v_I \gtrsim v_T$, produces an ultrashort THz pulse with an off-axis intensity maximum resembling a Cherenkov shock. The flexibility afforded by these regimes can be used to tailor the properties of the THz radiation for use as either an ultrafast, quasi-DC pump or probe.

Figure 1 illustrates the concept of two-color terahertz generation for a  conventional focusing geometry (a) and the ultrafast flying focus configuration applied in this work (b). In both cases, an initial laser pulse with frequency $\omega$ passes through a nonlinear crystal (e.g., beta-barium borate), which frequency doubles a portion of the energy. The phase and polarization of the second harmonic ($2\omega$) are adjusted using the location of the crystal and a dichroic waveplate to optimize the THz signal. In the conventional-focusing geometry, the composite two-color pulse is focused into the gas by an ideal lens where it drives an ionization front with a velocity $v_I$. The velocity has a complex dependence on the dispersion and propagation history of the pulse, offering little to no control over the THz emission angle [Eq. \eqref{eq:phi}]. For typical parameters, $v_I > v_T$, resulting in off-axis, conical emission ($\phi>0$) \cite{johnson2013thz}.

\begin{figure}
\includegraphics[scale =0.95]{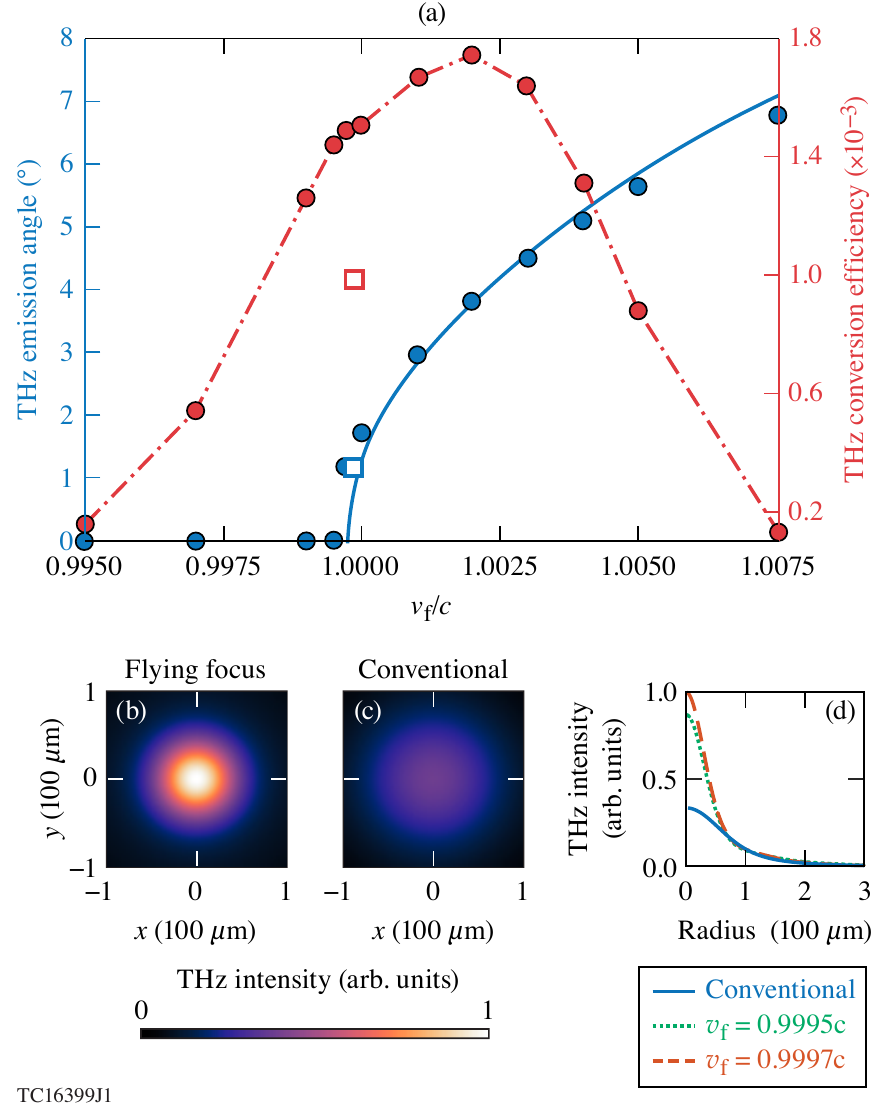}% Here is how to import EPS art
\caption{ \label{fig:f2} (a) The THz emission angle (solid blue) and conversion efficiency (dashed red) to frequencies $<30$ THz  as a function of focal velocity for the flying focus design (dots) and at the group velocity of the fundamental for the conventional design, $v_g =0.9997c$ (boxes). The flying focus design allows for on-axis THz generation with a conversion efficiency comparable to the conventional design. For focal velocities greater than the vacuum speed of light, the simulations are in agreement with the prediction of Eq.\eqref{eq:phi} (solid blue line). (b and c) The focused profile of the THz radiation at peak intensity for $v_f =0.9995c$ and the conventional pulse, respectively. (d) Lineouts of the focused intensity profile. The flying focus design produces highly focusable THz radiation.}
\end{figure}

Replacing the ideal lens with an axiparabola-echelon pair provides control over the velocity of the ionization front and THz emission angle (Fig. \ref{fig:f2}). The axiparabola focuses different radial locations in the near field to different axial locations in the far field, while the echelon adjusts the relative timing of the foci. The combination creates an ultrashort flying focus that travels at a velocity $v_f$ independent of the group velocity over distances much longer than a Rayleigh range. Because the ionization rate is sensitive to the intensity, the trajectory of the ionization front closely tracks that of the focus $v_I \approx v_f$. The ionization front driven by the flying focus can travel faster [Fig. \ref{fig:f1}(b) top], equal to [Fig. \ref{fig:f1}(b) middle], or slower than [Fig. \ref{fig:f1}(b) bottom] the THz phase velocity and generate radiation peaked either on or off-axis. 

The axiparabola extends the range of high intensity using  spherical and higher-order geometric aberration \cite{smartsev2019axiparabola,oubrerie2022axiparabola}. The focal length as a function of near-field radius $f(r')$ is designed to ensure that a flattop transverse intensity profile incident on the optic results in a uniform on-axis intensity in the far field. Specifically, $f(r')=f_0+(r'/R)^2L$, where $f_0$ is the nominal focal length, $R$ is the maximum radius, and the focal range $L$ determines the distance over which the flying focus maintains a high intensity. 

When used on its own, the axiparabola produces an accelerating focus. A specified, constant focal velocity $v_f$ can be achieved by applying a small amount of radial group delay (i.e., an intensity profile that depends non-separably on radius and time) as outlined in Refs. \cite{palastro2020dephasingless,oubrerie2022axiparabola,caizergues2020phase,ambat2023}. Here a reflective echelon is used for this purpose \cite{palastro2020dephasingless,oliver2020,pigeon2023}. The echelon consists of discrete, concentric rings with widths $\Delta r'$ determined by the desired radial group delay and depths $d$ equal to a half-integer multiple of the fundamental wavelength $d = (\ell/2)\lambda_1 = \pi\ell/\omega$. A single echelon can be used for both harmonics because their wavelengths are integer multiples, $\lambda_1 = 2 \lambda_2$. The widths of the rings are large enough to prevent diffractive losses: $\Delta r' \gg \lambda_1f_0/2R$. For $v_f \approx c$, the axiparabola-echelon system preserves the ultrashort optical pulse duration, making it ideal for efficient, ultrafast THz generation. 

To demonstrate spatiotemporal control of two-color THz generation, the interactions illustrated in Fig. \ref{fig:f1} were simulated using the unidirectional pulse propagation equation (UPPE) \cite{kolesik2002unidirectional,kolesik2004nonlinear}. The pulses were initialized with the phases and amplitudes produced by the optical configurations shown in Fig. \ref{fig:f1}. In addition to accounting for dispersion to all orders, the UPPE included source terms for the third-order bound electron nonlinearity (i.e., $\chi^{(3)}$), the free-electron current density, and the depletion of laser pulse energy (see Supplemental for details) \cite{johnson2013thz,kolesik2002unidirectional,kolesik2004nonlinear,couairon2011practitioner, babushkin2010ultrafast,berge20133d}. The parameters of the laser pulse (Table 1) were chosen to emulate the output of a commercially available multi-kHz Ti:Sapphire laser system \cite{KMLabs}. 

For the simulations that implemented the axiparabola-echelon pair, the focal range $L$ was approximately $150$ Rayleigh lengths of the full-aperture focal spot and roughly one half of the dephasing length between the two harmonics, $L_{\pi} \approx (\lambda_1/4) |n(\lambda_1)-n(\lambda_2)|^{-1}$, where $n$ is the refractive index. The profile of the echelon was varied to produce the desired focal velocity as calculated in Ref. \cite{palastro2020dephasingless}. The simulations of the conventional configuration used an ideal lens with an f-number $f/{\#} = 85$, so that the confocal parameter matched the focal range of the flying focus configuration, providing a comparable range of high intensity. In each configuration, the initial relative phase between the two harmonics was tuned to ensure the highest THz yield. During ionization the optimal relative phase for THz generation is $\pi/2$ \cite{kim2007terahertz}. However, the optimal initial phase may be different due to the evolution of the fundamental and second harmonic phases during propagation.

\begin{table}[b]
\caption{\label{tab:table1}%
Simulation parameters for the drive laser pulse, optical configurations, and gas. 
}
\begin{ruledtabular}
\begin{tabular}{lc}
\textrm{Laser Parameters}&
\textrm{Value}\\
\hline
$\lambda_1 (\mu \mathrm{m}$) & $0.8$ \\
$\lambda_1$ Power (GW) & 5\\
$\lambda_1$ Duration (fs) & 30\\
$\lambda_2 (\mu \mathrm{m})$ & $0.4$ \\
$\lambda_2$ Power (GW) & 1\\
$\lambda_2$ Duration (fs) & $15\sqrt{2}$\\
Energy (mJ) & 0.3\\
Supergaussian radial order & 20\\
Supergaussian temporal order & 2\\
\hline
\textrm{Optics Parameters}&
\textrm{Value}\\
\hline
Axiparabola radius $R$ (cm) & 5\\
Axiparabola $f/\#$ & 10\\
Focal range $L$ (cm) & 1.5\\
\hline
Ideal lens radius (cm) & 2.5\\
Ideal lens $f/\#$ & 85\\
Confocal parameter (cm) & 1.5\\
\hline
\textrm{Medium Parameters}&
\textrm{Value}\\
\hline
Species & Ar\\
Density ($\mathrm{cm}^{-3}$) & $2.7{\times}10^{19}$\\
Nonlinear refractive index ($ \mathrm{cm}^2/\mathrm{W}$) & $1{\times}10^{-19}$ \cite{zahedpour2015measurement}\\
Ionization energy (eV) & 15.8\\
Collision frequency (THz) & 10 \cite{sprangle2004ultrashort}\\

\end{tabular}
\end{ruledtabular}
\end{table}

The simulation results presented in Fig. \ref{fig:f2}(a) highlight the control over the THz emission angle enabled by the two-color flying focus. For focal velocities less than the THz phase velocity \cite{peck1964dispersion}, the THz is emitted on-axis, i.e., with $\phi = 0$, generating a directed, high-power THz pulse. For focal velocities greater than the THz phase velocity, the emission angles agree with the Cherenkov model [Eq. \eqref{eq:phi}] evaluated using $v_I = v_f$. The agreement confirms that the ionization front driven by the flying focus pulse has a velocity $v_I \approx v_f$. The conventional configuration produces off-axis THz emission with $\phi \approx 1^{\circ}$. In all cases, the emission angle is determined by propagating the THz radiation to the near field, time-integrating, and finding the radius of the peak fluence.

The flying focus yields larger THz conversion efficiencies than the conventional pulse for emission angles ranging from $\phi = 0$ to $5^{\circ}$. The elevated conversion efficiency at moderately superluminal focal velocities ($v_f \gtrsim c)$ is explained by three effects. First, a superluminal focus mitigates the ionization refraction experienced by the two-color pulse \cite{palastro2018ionization, turnbull2018ionization, simpson2020nonlinear}, resulting in a slightly higher peak intensity and a stronger THz-generating current. Second, THz radiation emitted at a larger angle spends less time within the plasma and thus undergoes less collisional absorption. Finally, as shown later in Fig. 4, superluminal focal velocities produce more THz radiation at higher frequencies where collisional absorption is weaker. Below $v_f\approx 0.998 \,c$ and beyond $v_f\approx1.002 \,c$, the flying focus optical assembly causes a drop in the conversion efficiency. As the focal velocity gets further from the group velocity, the echelon must impart a larger radial group delay. The associated chromatic aberration lengthens the effective duration of the moving focus and decreases its maximum intensity, both of which inhibit the THz generation.

Figures \ref{fig:f2}(b)-(d) demonstrate that the flying focus drives highly focusable THz radiation. The images in (b) and (c) show the maximally focused profile for all frequencies $\lesssim 30$ THz from the $v_f = 0.9995\,c$ ($\phi=0^{\circ}$) flying focus and conventional ($\phi=1^{\circ}$) configurations, respectively. The THz radiation  produced by the flying focus has a smaller spot and a higher intensity: the $1/e^2$ radius is $85\mu \mathrm{m}$ compared to $140\mu \mathrm{m}$ in the conventional case.  As illustrated by the lineouts in Fig. \ref{fig:f2}(d), this is not a result of the larger angle in the conventional case and is a general property of the THz generated by the flying focus. The profiles in Figs. \ref{fig:f2}(b)-(d) were obtained by propagating the radiation backwards in space excluding all source terms and dispersion to find the location and time of highest intensity. This is equivalent to finding the ``virtual'' THz source.  The increased focusability of THz radiation driven by the flying focus may enhance applications that require a high peak intensity or a small spot.

\begin{figure}[t]
\includegraphics[scale=1]{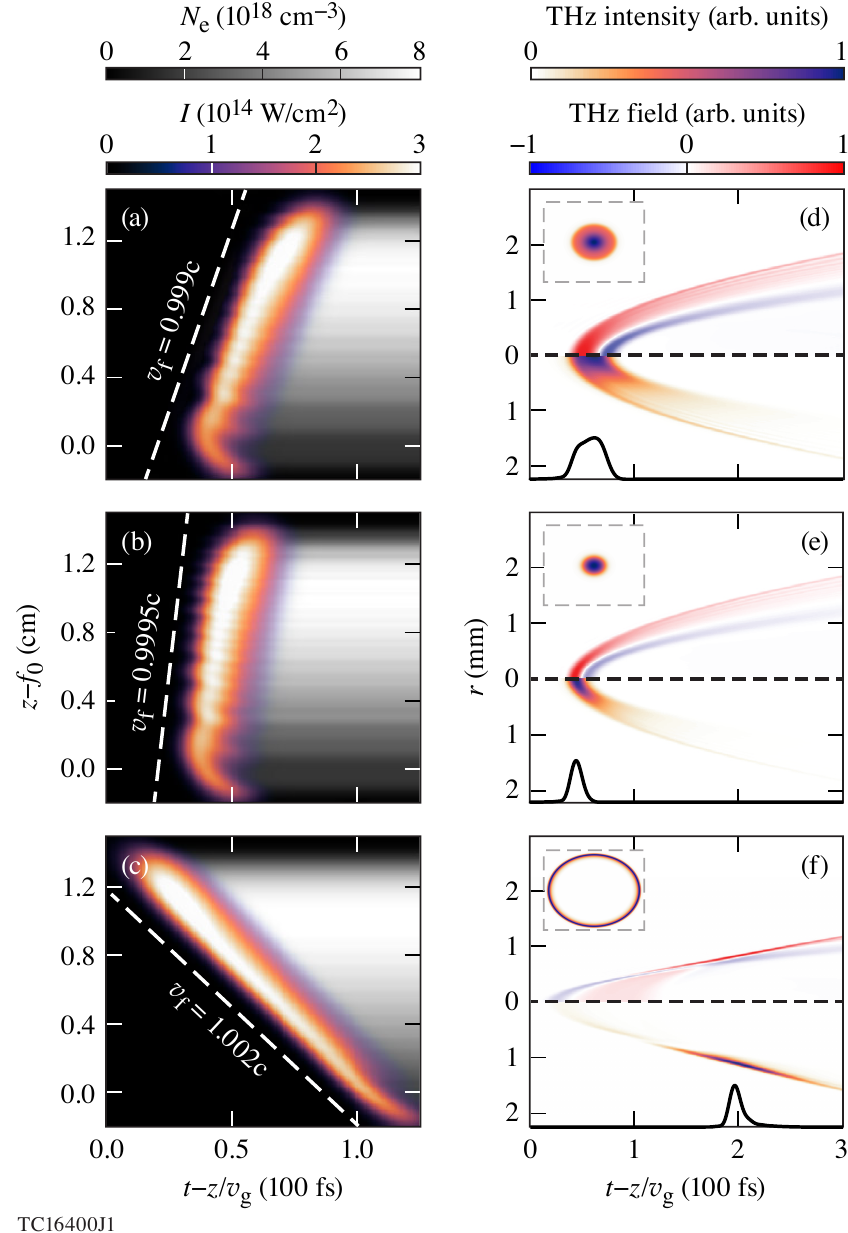}% Here is how to import EPS art
\caption{\label{fig:f3} (a)-(c) The cycle-averaged, on-axis intensity of the two-color flying focus pulse overlayed on the electron density as a function of distance and time in the moving frame for $v_f=0.999\,c$, $v_f=0.9995\,c$, and $v_f = 1.002\,c$, respectively. The dashed white lines indicate the designed focal trajectory in the moving frame coordinate $t-z/v_g$, where $v_g$ is the group velocity of the first harmonic. In this frame, the location of the focus $z_f = v_f(1-v_f/v_g)^{-1}(t-z/v_g)$. (d)-(f) The corresponding electric fields (top) and intensities (bottom) of all low frequency radiation ($\leq100\, \mathrm{THz}$) at the end of the focal range, i.e., at $z-f_0=1.5\,\mathrm{cm}$. The lineouts show the temporal profile at the radial location of the peak intensity, while the insets show the transverse profile. The focal velocity determines the spatiotemporal profile of the THz radiation in the far field.}
\end{figure}

In addition to the emission angle, the velocity of the ionization front determines the far-field spatiotemporal profile of the THz radiation. Figures \ref{fig:f3}(a)-(c) display the on-axis intensity of the two-color pulse and the resulting ionization front for $v_f\lesssim v_T$, $v_f \approx v_T$, and $v_f\gtrsim v_T$, respectively. Figures \ref{fig:f3}(d)-(f) show the final far-field THz waveform for each case in terms of the electric field (top) and intensity (bottom). In all three cases, the ionization front travels at $v_I \approx v_f$ and produces a single-cycle THz pulse with a distinct spatiotemporal profile. When $v_I\lesssim v_T$, the ionization front continually slips behind the radiation that was generated upstream and drives new radiation. This produces a longer THz pulse with an on-axis intensity maximum  [Figs. \ref{fig:f3}(a) \& (d)]. When $v_I\approx v_T$, the newly driven THz radiation overlaps the radiation generated upstream. The accumulation produces a shorter-duration THz pulse, also with an on-axis intensity maximum [Figs. \ref{fig:f3}(b) \& (e)]. When $v_I\gtrsim v_T$, the ionization front outpaces and drives new radiation ahead of the radiation generated upstream. The receding wavefronts constructively interfere at the Cherenkov angle, producing the shortest THz pulse, but with an off-axis intensity maximum [\ref{fig:f3}(c) \& (f)].  

\begin{figure}
\includegraphics[scale = 0.8]{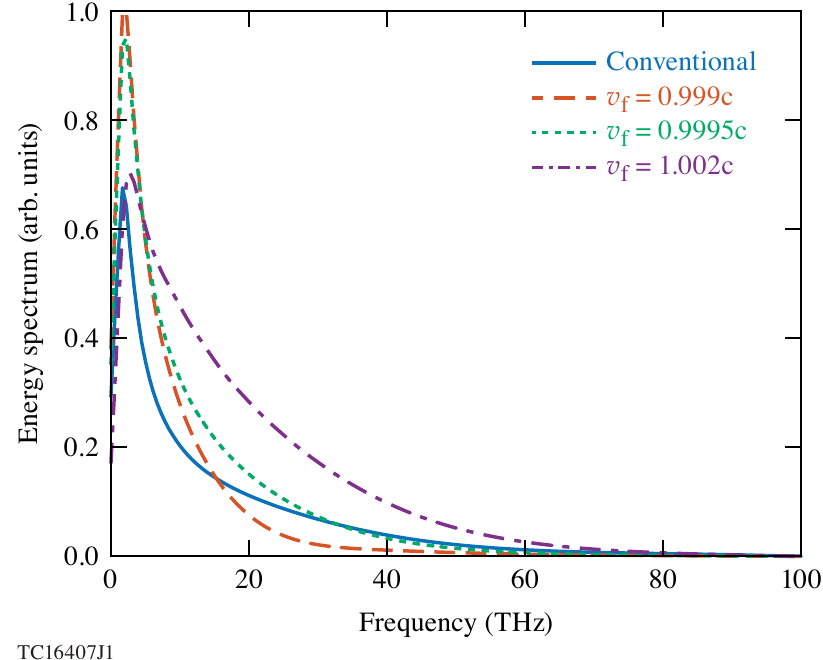}% Here is how to import EPS art
\caption{ \label{fig:f4} The energy spectrum of all generated low-frequency radiation ($\leq 100 \,\mathrm{THz}$) for the three focal velocities presented in Fig. \ref{fig:f3} and the THz pulse produced by the conventional geometry.}
\end{figure}

The change in the THz pulse duration with focal velocity (and ionizaton front velocity) is reflected in the spectrum. Figure \ref{fig:f4} shows the energy spectrum of all generated low frequency radiation, i.e., $\leq 100$ THz, for the three cases displayed in Fig. \ref{fig:f3} and the conventional focusing geometry. A faster ionization front results in a broader THz bandwidth, consistent with the shorter durations observed in Fig. \ref{fig:f3}. The spectral width depends primarily on the relative velocities of the ionization front and THz phase fronts and is not a consequence of the flying focus focal geometry. Simulations (not shown) which artificially altered the THz phase velocity by modifying the dispersion (similar to Ref. \cite{johnson2013thz}) produced qualitatively similar results for the same value of $v_T/v_I$: the THz spectrum was broader when $v_T<v_I$ and narrower when $v_T>v_I$. 

The simulations modeled THz generation over less than one dephasing length between the two harmonics $L_{\pi}$. When the generation occurs over many dephasing lengths, Eq. \eqref{eq:phi} is modified as follows \cite{you2012off,johnson2013thz}: $\cos(\phi) = v_T(1/v_I \pm \pi/L_{\pi}\omega_T)$, where $\omega_T$ is the THz frequency. In this case, the emission angle can still be controlled by using the flying focus to vary the ionization front velocity. This will be a topic of future work.

The ultrafast flying focus provides unprecedented control over two-color THz generation. The programmable-velocity intensity peak drives an ultrashort ionization front over many Rayleigh ranges, producing a highly focusable, single-cycle THz pulse. By tuning the velocity of the intensity peak, the emission angle, spatiotemporal profile, and spectrum of the THz radiation can be tailored to a particular application, such as probing picosecond dynamics or pumping new states of matter. The tunability is enabled by an experimentally demonstrated optical configuration consisting of an axiparabola and radial echelon \cite{pigeon2023}. The configuration avoids the need for multiple intersecting laser pulses \cite{lee2023intense}, a structured plasma \cite{antonsen2007excitation}, or an external magnetic field \cite{yoshii1997radiation}. Combining the ultrafast flying focus with a mid-infrared ultrashort pulse laser \cite{koulouklidis2020observation} could provide a path to focused THz pulses approaching relativistic intensities (${\sim} 2\times10^{13}\,\mathrm{W/cm^2}$). Combining the design with an ultra-high-rep-rate laser could provide directed and controlled THz pulses at $500\,\mathrm{kHz}$, approaching watt-level average powers \cite{buldt2021gas}.  

\begin{acknowledgements}
The authors would like to thank R. Boni for productive discussions. 
\vspace{12pt}

This report was prepared as an account of work sponsored by an agency of the U.S. Government. Neither the U.S. Government nor any agency thereof, nor any of their employees, makes any warranty, express or implied, or assumes any legal liability or responsibility for the accuracy, completeness, or usefulness of any information, apparatus, product, or process disclosed, or represents that its use would not infringe privately owned rights. Reference herein to any specific commercial product, process, or service by trade name, trademark, manufacturer, or otherwise does not necessarily constitute or imply its endorsement, recommendation, or favoring by the U.S. Government or any agency thereof. The views and opinions of authors expressed herein do not necessarily state or reflect those of the U.S. Government or any agency thereof.

This material is based upon work supported by the Office of Fusion Energy Sciences under Award Number DE-SC00215057, the Department of Energy National Nuclear Security Administration under Award Number DE-NA0003856, the University of Rochester, and the New York State Energy Research and Development Authority.

\end{acknowledgements}

\newpage
\begin{center}
\textbf{\large Supplemental: Spatiotemporal control of two-color terahertz generation}
\end{center}
%%%%%%%%%% Merge with supplemental materials %%%%%%%%%%
%%%%%%%%%% Prefix a "S" to all equations, figures, tables and reset the counter %%%%%%%%%%
\setcounter{equation}{0}
\setcounter{figure}{0}
\setcounter{table}{0}
\setcounter{page}{1}
\makeatletter
\renewcommand{\theequation}{S\arabic{equation}}
\renewcommand{\thefigure}{S\arabic{figure}}

The simulations of two-color THz generation were performed in two steps. The first step used the frequency domain Fresnel integral to propagate the two-color laser pulse from the optical assembly to the far field. The second step used the unidirectional pulse propagation equation (UPPE) \cite{kolesik2002unidirectional,kolesik2004nonlinear,couairon2011practitioner} to propagate the pulse through the far field and to model the generation and propagation of the THz radiation. 

The transverse electric field of the two-color laser pulse was initialized just before the optical assembly with the parameters provided in Table 1. The phase imparted by the optical assembly, i.e., either an axiparabola-echelon pair or the ideal lens, was then applied to the initial field. After the optical assembly at $z=0$, the field can be expressed as $\tilde{E}_0(r',\omega)e^{i\theta(r',\omega)}$, where  $\tilde{E}_0(r',\omega)$ is the field before the assembly, $\theta(r',\omega)$ is the phase applied by the assembly, $r'$ is the near-field radius, and $\sim$ indicates a frequency-domain, as opposed to a time-domain, field. The field was propagated in vacuum from $z=0$ to the beginning of the gas at $z=L$ using the frequency-domain Fresnel integral:
\begin{equation}
\begin{split}
    \tilde{E}(r,\omega,z=L) = 
    \frac{\omega}{ic L}e^{-i\omega L/c
    }\int r' dr' \tilde{E}_0(r',\omega) \times ... \\
    J_0\bigg(\frac{\omega r r'}{cL}\bigg)
    \exp{\bigg[\frac{i\omega(r^2+r'^2)}{2cL}+i\theta(r',\omega)\bigg]},
\end{split}
\end{equation}
where $r$ is the far-field radius and $J_0$ is the zeroth-order Bessel function of the first kind. The Fresnel integral decouples the radial grids in the near field and far field, reducing computational expense, especially when considering smaller $f/\#$'s \cite{palastro2018ionization}. 

The electric field from the Fresnel integral $\tilde{E}(r,\omega,z=L)$ provided the initial condition for the UPPE. The UPPE models linear dispersion to all orders, does not make a paraxial approximation, and can evolve an arbitrarily broad frequency spectrum.  As implemented here, the UPPE is expressed as
\begin{equation}
        \frac{\partial\hat{E}}{\partial z}=i\left(k_z-\frac{\omega}{v_{g}}\right)\hat{E}+\frac{\omega}{2\varepsilon_0c^2k_z}(i\omega \hat{P}_{g}-\hat{J}_{p}-\hat{J}_{i}),
\end{equation}
where $k_z= \sqrt{k^2_\omega-k^2_r}$ is the $z$-component of the wavevector, $k_{\omega} = n(\omega)\omega/c$, $n(\omega)$ is the refractive index, $k_r$ is the transverse wavevector, $v_g$ is the group velocity at the fundamental frequency of the laser pulse, and \^{} indicates a Fourier and Hankel transformed quantity. The source terms correspond to the nonlinear polarization density of the gas $\hat{P}_{g}$, the current density of the plasma $\hat{J}_{p}$, and an effective current density that accounts for the electromagnetic energy lost during photoionization $\hat{J}_{i}$. Each of these are calculated in the time and configuration domain as follows:
\begin{align}
    P_{g} &= \frac{4}{3}\varepsilon^2_0cn_2E^3 \\
    \partial_t J_{p} &= -v_{en}J_{p}+\frac{e^2}{m_e}n_eE \\
    J_{i} &= \frac{w(E)n_nU_I}{E},
\end{align}
where $n_2$ is the nonlinear refractive index, $v_{en}$ is the electron-neutral collision frequency, $n_e$ is the electron density, $w(E)$ is the ionization rate, $n_n$ is the neutral density, and $U_I$ is the ionization energy. The electron density and neutral densities evolve according to $\partial_t n_e = w(E)n_n$ and $n_n = n_0 - n_e$, where $n_0$ is the initial neutral density. 

The UPPE model is solved in a cylindrically symmetric 2D$+t$ geometry using quasi-discrete Hankel transforms \cite{guizar2004computation} with a second-order predictor-corrector scheme for the nonlinear source terms. The Ammosov-Delone-Krainov (ADK) \cite{ammosov1986tunnel} rate is used for $w(E)$, and a fixed electron-neutral collision frequency is implemented in the current density equation to account for inverse bremsstrahlung absorption \cite{richardson20192019}. The background gas was singly-ionizable argon, with parameters given in Table 1. The refractive index was calculated using the Sellmeier equation given in Ref. \cite{peck1964dispersion} and extrapolated to THz frequencies. 

The temporal resolution and domain sizes were $\Delta t=110 \,\mathrm{as}$ and $T = 1.8\,\mathrm{ps}$, corresponding to a spectral resolution of $\Delta f= 0.55 \, \mathrm{THz}$. The minimum radial resolution was $\Delta r = 1.5\,\mu\mathrm{m}$ with a radial domain size $r_{\mathrm{max}} = 4.5\,\mathrm{mm}$. The axial step size was $\Delta z = 20 \,\mu\mathrm{m}$. The resolution was adjusted to ensure convergence, and the maximum bounds on $t$ and $r$ were chosen to mitigate aliasing and spurious reflections, respectively.

\newpage

\bibliography{bibliography}

\providecommand{\noopsort}[1]{}\providecommand{\singleletter}[1]{#1}%
\begin{thebibliography}{59}%
\makeatletter
\providecommand \@ifxundefined [1]{%
 \@ifx{#1\undefined}
}%
\providecommand \@ifnum [1]{%
 \ifnum #1\expandafter \@firstoftwo
 \else \expandafter \@secondoftwo
 \fi
}%
\providecommand \@ifx [1]{%
 \ifx #1\expandafter \@firstoftwo
 \else \expandafter \@secondoftwo
 \fi
}%
\providecommand \natexlab [1]{#1}%
\providecommand \enquote  [1]{``#1''}%
\providecommand \bibnamefont  [1]{#1}%
\providecommand \bibfnamefont [1]{#1}%
\providecommand \citenamefont [1]{#1}%
\providecommand \href@noop [0]{\@secondoftwo}%
\providecommand \href [0]{\begingroup \@sanitize@url \@href}%
\providecommand \@href[1]{\@@startlink{#1}\@@href}%
\providecommand \@@href[1]{\endgroup#1\@@endlink}%
\providecommand \@sanitize@url [0]{\catcode `\\12\catcode `\$12\catcode
  `\&12\catcode `\#12\catcode `\^12\catcode `\_12\catcode `\%12\relax}%
\providecommand \@@startlink[1]{}%
\providecommand \@@endlink[0]{}%
\providecommand \url  [0]{\begingroup\@sanitize@url \@url }%
\providecommand \@url [1]{\endgroup\@href {#1}{\urlprefix }}%
\providecommand \urlprefix  [0]{URL }%
\providecommand \Eprint [0]{\href }%
\providecommand \doibase [0]{https://doi.org/}%
\providecommand \selectlanguage [0]{\@gobble}%
\providecommand \bibinfo  [0]{\@secondoftwo}%
\providecommand \bibfield  [0]{\@secondoftwo}%
\providecommand \translation [1]{[#1]}%
\providecommand \BibitemOpen [0]{}%
\providecommand \bibitemStop [0]{}%
\providecommand \bibitemNoStop [0]{.\EOS\space}%
\providecommand \EOS [0]{\spacefactor3000\relax}%
\providecommand \BibitemShut  [1]{\csname bibitem#1\endcsname}%
\let\auto@bib@innerbib\@empty
%</preamble>
\bibitem [{\citenamefont {Ulbricht}\ \emph {et~al.}(2011)\citenamefont
  {Ulbricht}, \citenamefont {Hendry}, \citenamefont {Shan}, \citenamefont
  {Heinz},\ and\ \citenamefont {Bonn}}]{ulbricht2011carrier}%
  \BibitemOpen
  \bibfield  {author} {\bibinfo {author} {\bibfnamefont {R.}~\bibnamefont
  {Ulbricht}}, \bibinfo {author} {\bibfnamefont {E.}~\bibnamefont {Hendry}},
  \bibinfo {author} {\bibfnamefont {J.}~\bibnamefont {Shan}}, \bibinfo {author}
  {\bibfnamefont {T.~F.}\ \bibnamefont {Heinz}},\ and\ \bibinfo {author}
  {\bibfnamefont {M.}~\bibnamefont {Bonn}},\ }\bibfield  {title} {\bibinfo
  {title} {Carrier dynamics in semiconductors studied with time-resolved
  terahertz spectroscopy},\ }\href@noop {} {\bibfield  {journal} {\bibinfo
  {journal} {Reviews of Modern Physics}\ }\textbf {\bibinfo {volume} {83}},\
  \bibinfo {pages} {543} (\bibinfo {year} {2011})}\BibitemShut {NoStop}%
\bibitem [{\citenamefont {Li}\ \emph {et~al.}(2022)\citenamefont {Li},
  \citenamefont {Medapalli}, \citenamefont {Mentink}, \citenamefont
  {Mikhaylovskiy}, \citenamefont {Blank}, \citenamefont {Patel}, \citenamefont
  {Zvezdin}, \citenamefont {Rasing}, \citenamefont {Fullerton},\ and\
  \citenamefont {Kimel}}]{li2022ultrafast}%
  \BibitemOpen
  \bibfield  {author} {\bibinfo {author} {\bibfnamefont {G.}~\bibnamefont
  {Li}}, \bibinfo {author} {\bibfnamefont {R.}~\bibnamefont {Medapalli}},
  \bibinfo {author} {\bibfnamefont {J.~H.}\ \bibnamefont {Mentink}}, \bibinfo
  {author} {\bibfnamefont {R.~V.}\ \bibnamefont {Mikhaylovskiy}}, \bibinfo
  {author} {\bibfnamefont {T.}~\bibnamefont {Blank}}, \bibinfo {author}
  {\bibfnamefont {S.~K.~K.}\ \bibnamefont {Patel}}, \bibinfo {author}
  {\bibfnamefont {A.~K.}\ \bibnamefont {Zvezdin}}, \bibinfo {author}
  {\bibfnamefont {T.~H.}\ \bibnamefont {Rasing}}, \bibinfo {author}
  {\bibfnamefont {E.~E.}\ \bibnamefont {Fullerton}},\ and\ \bibinfo {author}
  {\bibfnamefont {A.~V.}\ \bibnamefont {Kimel}},\ }\bibfield  {title} {\bibinfo
  {title} {Ultrafast kinetics of the antiferromagnetic-ferromagnetic phase
  transition in ferh},\ }\href@noop {} {\bibfield  {journal} {\bibinfo
  {journal} {Nature Communications}\ }\textbf {\bibinfo {volume} {13}},\
  \bibinfo {pages} {2998} (\bibinfo {year} {2022})}\BibitemShut {NoStop}%
\bibitem [{\citenamefont {Chen}\ \emph {et~al.}(2021)\citenamefont {Chen},
  \citenamefont {Curry}, \citenamefont {Zhang}, \citenamefont {Treffert},
  \citenamefont {Stojanovic}, \citenamefont {Toleikis}, \citenamefont {Pan},
  \citenamefont {Gauthier}, \citenamefont {Zapolnova}, \citenamefont {Seipp}
  \emph {et~al.}}]{chen2021ultrafast}%
  \BibitemOpen
  \bibfield  {author} {\bibinfo {author} {\bibfnamefont {Z.}~\bibnamefont
  {Chen}}, \bibinfo {author} {\bibfnamefont {C.~B.}\ \bibnamefont {Curry}},
  \bibinfo {author} {\bibfnamefont {R.}~\bibnamefont {Zhang}}, \bibinfo
  {author} {\bibfnamefont {F.}~\bibnamefont {Treffert}}, \bibinfo {author}
  {\bibfnamefont {N.}~\bibnamefont {Stojanovic}}, \bibinfo {author}
  {\bibfnamefont {S.}~\bibnamefont {Toleikis}}, \bibinfo {author}
  {\bibfnamefont {R.}~\bibnamefont {Pan}}, \bibinfo {author} {\bibfnamefont
  {M.}~\bibnamefont {Gauthier}}, \bibinfo {author} {\bibfnamefont
  {E.}~\bibnamefont {Zapolnova}}, \bibinfo {author} {\bibfnamefont {L.~E.}\
  \bibnamefont {Seipp}}, \emph {et~al.},\ }\bibfield  {title} {\bibinfo {title}
  {Ultrafast multi-cycle terahertz measurements of the electrical conductivity
  in strongly excited solids},\ }\href@noop {} {\bibfield  {journal} {\bibinfo
  {journal} {Nature Communications}\ }\textbf {\bibinfo {volume} {12}},\
  \bibinfo {pages} {1638} (\bibinfo {year} {2021})}\BibitemShut {NoStop}%
\bibitem [{\citenamefont {Markelz}\ and\ \citenamefont
  {Mittleman}(2022)}]{markelz2022perspective}%
  \BibitemOpen
  \bibfield  {author} {\bibinfo {author} {\bibfnamefont {A.~G.}\ \bibnamefont
  {Markelz}}\ and\ \bibinfo {author} {\bibfnamefont {D.~M.}\ \bibnamefont
  {Mittleman}},\ }\bibfield  {title} {\bibinfo {title} {Perspective on
  terahertz applications in bioscience and biotechnology},\ }\href@noop {}
  {\bibfield  {journal} {\bibinfo  {journal} {ACS Photonics}\ }\textbf
  {\bibinfo {volume} {9}},\ \bibinfo {pages} {1117} (\bibinfo {year}
  {2022})}\BibitemShut {NoStop}%
\bibitem [{\citenamefont {Disa}\ \emph {et~al.}(2023)\citenamefont {Disa},
  \citenamefont {Curtis}, \citenamefont {Fechner}, \citenamefont {Liu},
  \citenamefont {von Hoegen}, \citenamefont {F{\"o}rst}, \citenamefont {Nova},
  \citenamefont {Narang}, \citenamefont {Maljuk}, \citenamefont {Boris} \emph
  {et~al.}}]{disa2023photo}%
  \BibitemOpen
  \bibfield  {author} {\bibinfo {author} {\bibfnamefont {A.~S.}\ \bibnamefont
  {Disa}}, \bibinfo {author} {\bibfnamefont {J.}~\bibnamefont {Curtis}},
  \bibinfo {author} {\bibfnamefont {M.}~\bibnamefont {Fechner}}, \bibinfo
  {author} {\bibfnamefont {A.}~\bibnamefont {Liu}}, \bibinfo {author}
  {\bibfnamefont {A.}~\bibnamefont {von Hoegen}}, \bibinfo {author}
  {\bibfnamefont {M.}~\bibnamefont {F{\"o}rst}}, \bibinfo {author}
  {\bibfnamefont {T.~F.}\ \bibnamefont {Nova}}, \bibinfo {author}
  {\bibfnamefont {P.}~\bibnamefont {Narang}}, \bibinfo {author} {\bibfnamefont
  {A.}~\bibnamefont {Maljuk}}, \bibinfo {author} {\bibfnamefont {A.~V.}\
  \bibnamefont {Boris}}, \emph {et~al.},\ }\bibfield  {title} {\bibinfo {title}
  {Photo-induced high-temperature ferromagnetism in ytio3},\ }\href@noop {}
  {\bibfield  {journal} {\bibinfo  {journal} {Nature}\ }\textbf {\bibinfo
  {volume} {617}},\ \bibinfo {pages} {73} (\bibinfo {year} {2023})}\BibitemShut
  {NoStop}%
\bibitem [{\citenamefont {F{\"u}l{\"o}p}\ \emph {et~al.}(2020)\citenamefont
  {F{\"u}l{\"o}p}, \citenamefont {Tzortzakis},\ and\ \citenamefont
  {Kampfrath}}]{fulop2020laser}%
  \BibitemOpen
  \bibfield  {author} {\bibinfo {author} {\bibfnamefont {J.~A.}\ \bibnamefont
  {F{\"u}l{\"o}p}}, \bibinfo {author} {\bibfnamefont {S.}~\bibnamefont
  {Tzortzakis}},\ and\ \bibinfo {author} {\bibfnamefont {T.}~\bibnamefont
  {Kampfrath}},\ }\bibfield  {title} {\bibinfo {title} {Laser-driven
  strong-field terahertz sources},\ }\href@noop {} {\bibfield  {journal}
  {\bibinfo  {journal} {Advanced Optical Materials}\ }\textbf {\bibinfo
  {volume} {8}},\ \bibinfo {pages} {1900681} (\bibinfo {year}
  {2020})}\BibitemShut {NoStop}%
\bibitem [{\citenamefont {Lewis}(2014)}]{lewis2014review}%
  \BibitemOpen
  \bibfield  {author} {\bibinfo {author} {\bibfnamefont {R.~A.}\ \bibnamefont
  {Lewis}},\ }\bibfield  {title} {\bibinfo {title} {A review of terahertz
  sources},\ }\href@noop {} {\bibfield  {journal} {\bibinfo  {journal} {Journal
  of Physics D: Applied Physics}\ }\textbf {\bibinfo {volume} {47}},\ \bibinfo
  {pages} {374001} (\bibinfo {year} {2014})}\BibitemShut {NoStop}%
\bibitem [{\citenamefont {Leemans}\ \emph {et~al.}(2003)\citenamefont
  {Leemans}, \citenamefont {Geddes}, \citenamefont {Faure}, \citenamefont
  {T\'oth}, \citenamefont {van Tilborg}, \citenamefont {Schroeder},
  \citenamefont {Esarey}, \citenamefont {Fubiani}, \citenamefont {Auerbach},
  \citenamefont {Marcelis}, \citenamefont {Carnahan}, \citenamefont {Kaindl},
  \citenamefont {Byrd},\ and\ \citenamefont {Martin}}]{LeemansCTR2003}%
  \BibitemOpen
  \bibfield  {author} {\bibinfo {author} {\bibfnamefont {W.~P.}\ \bibnamefont
  {Leemans}}, \bibinfo {author} {\bibfnamefont {C.~G.~R.}\ \bibnamefont
  {Geddes}}, \bibinfo {author} {\bibfnamefont {J.}~\bibnamefont {Faure}},
  \bibinfo {author} {\bibfnamefont {C.}~\bibnamefont {T\'oth}}, \bibinfo
  {author} {\bibfnamefont {J.}~\bibnamefont {van Tilborg}}, \bibinfo {author}
  {\bibfnamefont {C.~B.}\ \bibnamefont {Schroeder}}, \bibinfo {author}
  {\bibfnamefont {E.}~\bibnamefont {Esarey}}, \bibinfo {author} {\bibfnamefont
  {G.}~\bibnamefont {Fubiani}}, \bibinfo {author} {\bibfnamefont
  {D.}~\bibnamefont {Auerbach}}, \bibinfo {author} {\bibfnamefont
  {B.}~\bibnamefont {Marcelis}}, \bibinfo {author} {\bibfnamefont {M.~A.}\
  \bibnamefont {Carnahan}}, \bibinfo {author} {\bibfnamefont {R.~A.}\
  \bibnamefont {Kaindl}}, \bibinfo {author} {\bibfnamefont {J.}~\bibnamefont
  {Byrd}},\ and\ \bibinfo {author} {\bibfnamefont {M.~C.}\ \bibnamefont
  {Martin}},\ }\bibfield  {title} {\bibinfo {title} {Observation of terahertz
  emission from a laser-plasma accelerated electron bunch crossing a
  plasma-vacuum boundary},\ }\href
  {https://doi.org/10.1103/PhysRevLett.91.074802} {\bibfield  {journal}
  {\bibinfo  {journal} {Phys. Rev. Lett.}\ }\textbf {\bibinfo {volume} {91}},\
  \bibinfo {pages} {074802} (\bibinfo {year} {2003})}\BibitemShut {NoStop}%
\bibitem [{\citenamefont {Kim}\ \emph {et~al.}(2007)\citenamefont {Kim},
  \citenamefont {Glownia}, \citenamefont {Taylor},\ and\ \citenamefont
  {Rodriguez}}]{kim2007terahertz}%
  \BibitemOpen
  \bibfield  {author} {\bibinfo {author} {\bibfnamefont {K.-Y.}\ \bibnamefont
  {Kim}}, \bibinfo {author} {\bibfnamefont {J.~H.}\ \bibnamefont {Glownia}},
  \bibinfo {author} {\bibfnamefont {A.~J.}\ \bibnamefont {Taylor}},\ and\
  \bibinfo {author} {\bibfnamefont {G.}~\bibnamefont {Rodriguez}},\ }\bibfield
  {title} {\bibinfo {title} {Terahertz emission from ultrafast ionizing air in
  symmetry-broken laser fields},\ }\href@noop {} {\bibfield  {journal}
  {\bibinfo  {journal} {Optics Express}\ }\textbf {\bibinfo {volume} {15}},\
  \bibinfo {pages} {4577} (\bibinfo {year} {2007})}\BibitemShut {NoStop}%
\bibitem [{\citenamefont {Antonsen}\ \emph {et~al.}(2007)\citenamefont
  {Antonsen}, \citenamefont {Palastro},\ and\ \citenamefont
  {Milchberg}}]{antonsen2007excitation}%
  \BibitemOpen
  \bibfield  {author} {\bibinfo {author} {\bibfnamefont {T.~M.}\ \bibnamefont
  {Antonsen}}, \bibinfo {author} {\bibfnamefont {J.~P.}\ \bibnamefont
  {Palastro}},\ and\ \bibinfo {author} {\bibfnamefont {H.~M.}\ \bibnamefont
  {Milchberg}},\ }\bibfield  {title} {\bibinfo {title} {Excitation of terahertz
  radiation by laser pulses in nonuniform plasma channels},\ }\href@noop {}
  {\bibfield  {journal} {\bibinfo  {journal} {Physics of plasmas}\ }\textbf
  {\bibinfo {volume} {14}} (\bibinfo {year} {2007})}\BibitemShut {NoStop}%
\bibitem [{\citenamefont {Albert}\ and\ \citenamefont
  {Thomas}(2016)}]{Albert_2016}%
  \BibitemOpen
  \bibfield  {author} {\bibinfo {author} {\bibfnamefont {F.}~\bibnamefont
  {Albert}}\ and\ \bibinfo {author} {\bibfnamefont {A.~G.~R.}\ \bibnamefont
  {Thomas}},\ }\bibfield  {title} {\bibinfo {title} {Applications of laser
  wakefield accelerator-based light sources},\ }\href
  {https://doi.org/10.1088/0741-3335/58/10/103001} {\bibfield  {journal}
  {\bibinfo  {journal} {Plasma Physics and Controlled Fusion}\ }\textbf
  {\bibinfo {volume} {58}},\ \bibinfo {pages} {103001} (\bibinfo {year}
  {2016})}\BibitemShut {NoStop}%
\bibitem [{\citenamefont {Hafez}\ \emph {et~al.}(2016)\citenamefont {Hafez},
  \citenamefont {Chai}, \citenamefont {Ibrahim}, \citenamefont {Mondal},
  \citenamefont {F{\'e}rachou}, \citenamefont {Ropagnol},\ and\ \citenamefont
  {Ozaki}}]{hafez2016intense}%
  \BibitemOpen
  \bibfield  {author} {\bibinfo {author} {\bibfnamefont {H.~A.}\ \bibnamefont
  {Hafez}}, \bibinfo {author} {\bibfnamefont {X.}~\bibnamefont {Chai}},
  \bibinfo {author} {\bibfnamefont {A.}~\bibnamefont {Ibrahim}}, \bibinfo
  {author} {\bibfnamefont {S.}~\bibnamefont {Mondal}}, \bibinfo {author}
  {\bibfnamefont {D.}~\bibnamefont {F{\'e}rachou}}, \bibinfo {author}
  {\bibfnamefont {X.}~\bibnamefont {Ropagnol}},\ and\ \bibinfo {author}
  {\bibfnamefont {T.}~\bibnamefont {Ozaki}},\ }\bibfield  {title} {\bibinfo
  {title} {Intense terahertz radiation and their applications},\ }\href@noop {}
  {\bibfield  {journal} {\bibinfo  {journal} {Journal of Optics}\ }\textbf
  {\bibinfo {volume} {18}},\ \bibinfo {pages} {093004} (\bibinfo {year}
  {2016})}\BibitemShut {NoStop}%
\bibitem [{\citenamefont {Miao}\ \emph {et~al.}(2017)\citenamefont {Miao},
  \citenamefont {Palastro},\ and\ \citenamefont {Antonsen}}]{miao2017high}%
  \BibitemOpen
  \bibfield  {author} {\bibinfo {author} {\bibfnamefont {C.}~\bibnamefont
  {Miao}}, \bibinfo {author} {\bibfnamefont {J.~P.}\ \bibnamefont {Palastro}},\
  and\ \bibinfo {author} {\bibfnamefont {T.~M.}\ \bibnamefont {Antonsen}},\
  }\bibfield  {title} {\bibinfo {title} {High-power tunable laser driven thz
  generation in corrugated plasma waveguides},\ }\href@noop {} {\bibfield
  {journal} {\bibinfo  {journal} {Physics of Plasmas}\ }\textbf {\bibinfo
  {volume} {24}},\ \bibinfo {pages} {043109} (\bibinfo {year}
  {2017})}\BibitemShut {NoStop}%
\bibitem [{\citenamefont {Liao}\ \emph {et~al.}(2020)\citenamefont {Liao},
  \citenamefont {Liu}, \citenamefont {Scott}, \citenamefont {Zhang},
  \citenamefont {Zhu}, \citenamefont {Zhang}, \citenamefont {Li}, \citenamefont
  {Armstrong}, \citenamefont {Zemaityte}, \citenamefont {Bradford},
  \citenamefont {Rusby}, \citenamefont {Neely}, \citenamefont {Huggard},
  \citenamefont {McKenna}, \citenamefont {Brenner}, \citenamefont {Woolsey},
  \citenamefont {Wang}, \citenamefont {Sheng},\ and\ \citenamefont
  {Zhang}}]{LiaoRelLaserFoil2020}%
  \BibitemOpen
  \bibfield  {author} {\bibinfo {author} {\bibfnamefont {G.-Q.}\ \bibnamefont
  {Liao}}, \bibinfo {author} {\bibfnamefont {H.}~\bibnamefont {Liu}}, \bibinfo
  {author} {\bibfnamefont {G.~G.}\ \bibnamefont {Scott}}, \bibinfo {author}
  {\bibfnamefont {Y.-H.}\ \bibnamefont {Zhang}}, \bibinfo {author}
  {\bibfnamefont {B.-J.}\ \bibnamefont {Zhu}}, \bibinfo {author} {\bibfnamefont
  {Z.}~\bibnamefont {Zhang}}, \bibinfo {author} {\bibfnamefont {Y.-T.}\
  \bibnamefont {Li}}, \bibinfo {author} {\bibfnamefont {C.}~\bibnamefont
  {Armstrong}}, \bibinfo {author} {\bibfnamefont {E.}~\bibnamefont
  {Zemaityte}}, \bibinfo {author} {\bibfnamefont {P.}~\bibnamefont {Bradford}},
  \bibinfo {author} {\bibfnamefont {D.~R.}\ \bibnamefont {Rusby}}, \bibinfo
  {author} {\bibfnamefont {D.}~\bibnamefont {Neely}}, \bibinfo {author}
  {\bibfnamefont {P.~G.}\ \bibnamefont {Huggard}}, \bibinfo {author}
  {\bibfnamefont {P.}~\bibnamefont {McKenna}}, \bibinfo {author} {\bibfnamefont
  {C.~M.}\ \bibnamefont {Brenner}}, \bibinfo {author} {\bibfnamefont {N.~C.}\
  \bibnamefont {Woolsey}}, \bibinfo {author} {\bibfnamefont {W.-M.}\
  \bibnamefont {Wang}}, \bibinfo {author} {\bibfnamefont {Z.-M.}\ \bibnamefont
  {Sheng}},\ and\ \bibinfo {author} {\bibfnamefont {J.}~\bibnamefont {Zhang}},\
  }\bibfield  {title} {\bibinfo {title} {Towards terawatt-scale spectrally
  tunable terahertz pulses via relativistic laser-foil interactions},\ }\href
  {https://doi.org/10.1103/PhysRevX.10.031062} {\bibfield  {journal} {\bibinfo
  {journal} {Phys. Rev. X}\ }\textbf {\bibinfo {volume} {10}},\ \bibinfo
  {pages} {031062} (\bibinfo {year} {2020})}\BibitemShut {NoStop}%
\bibitem [{\citenamefont {Dai}\ \emph {et~al.}(2009)\citenamefont {Dai},
  \citenamefont {Karpowicz},\ and\ \citenamefont {Zhang}}]{dai2009coherent}%
  \BibitemOpen
  \bibfield  {author} {\bibinfo {author} {\bibfnamefont {J.}~\bibnamefont
  {Dai}}, \bibinfo {author} {\bibfnamefont {N.}~\bibnamefont {Karpowicz}},\
  and\ \bibinfo {author} {\bibfnamefont {X.~C.}\ \bibnamefont {Zhang}},\
  }\bibfield  {title} {\bibinfo {title} {Coherent polarization control of
  terahertz waves generated from two-color laser-induced gas plasma},\
  }\href@noop {} {\bibfield  {journal} {\bibinfo  {journal} {Physical Review
  Letters}\ }\textbf {\bibinfo {volume} {103}},\ \bibinfo {pages} {023001}
  (\bibinfo {year} {2009})}\BibitemShut {NoStop}%
\bibitem [{\citenamefont {Babushkin}\ \emph {et~al.}(2010)\citenamefont
  {Babushkin}, \citenamefont {Kuehn}, \citenamefont {Koehler}, \citenamefont
  {Skupin}, \citenamefont {Berge}, \citenamefont {Reimann}, \citenamefont
  {Woerner}, \citenamefont {Herrmann},\ and\ \citenamefont
  {Elsaesser}}]{babushkin2010ultrafast}%
  \BibitemOpen
  \bibfield  {author} {\bibinfo {author} {\bibfnamefont {I.}~\bibnamefont
  {Babushkin}}, \bibinfo {author} {\bibfnamefont {W.}~\bibnamefont {Kuehn}},
  \bibinfo {author} {\bibfnamefont {C.}~\bibnamefont {Koehler}}, \bibinfo
  {author} {\bibfnamefont {S.}~\bibnamefont {Skupin}}, \bibinfo {author}
  {\bibfnamefont {L.}~\bibnamefont {Berge}}, \bibinfo {author} {\bibfnamefont
  {K.}~\bibnamefont {Reimann}}, \bibinfo {author} {\bibfnamefont
  {M.}~\bibnamefont {Woerner}}, \bibinfo {author} {\bibfnamefont
  {J.}~\bibnamefont {Herrmann}},\ and\ \bibinfo {author} {\bibfnamefont
  {T.}~\bibnamefont {Elsaesser}},\ }\bibfield  {title} {\bibinfo {title}
  {Ultrafast spatiotemporal dynamics of terahertz generation by ionizing
  two-color femtosecond pulses in gases},\ }\href@noop {} {\bibfield  {journal}
  {\bibinfo  {journal} {Physical Review Letters}\ }\textbf {\bibinfo {volume}
  {105}},\ \bibinfo {pages} {053903} (\bibinfo {year} {2010})}\BibitemShut
  {NoStop}%
\bibitem [{\citenamefont {Kim}\ \emph {et~al.}(2012)\citenamefont {Kim},
  \citenamefont {Glownia}, \citenamefont {Taylor},\ and\ \citenamefont
  {Rodriguez}}]{kim2012high}%
  \BibitemOpen
  \bibfield  {author} {\bibinfo {author} {\bibfnamefont {K.~Y.}\ \bibnamefont
  {Kim}}, \bibinfo {author} {\bibfnamefont {J.~H.}\ \bibnamefont {Glownia}},
  \bibinfo {author} {\bibfnamefont {A.~J.}\ \bibnamefont {Taylor}},\ and\
  \bibinfo {author} {\bibfnamefont {G.}~\bibnamefont {Rodriguez}},\ }\bibfield
  {title} {\bibinfo {title} {High-power broadband terahertz generation via
  two-color photoionization in gases},\ }\href@noop {} {\bibfield  {journal}
  {\bibinfo  {journal} {IEEE Journal of Quantum Electronics}\ }\textbf
  {\bibinfo {volume} {48}},\ \bibinfo {pages} {797} (\bibinfo {year}
  {2012})}\BibitemShut {NoStop}%
\bibitem [{\citenamefont {Clerici}\ \emph {et~al.}(2013)\citenamefont
  {Clerici}, \citenamefont {Peccianti}, \citenamefont {Schmidt}, \citenamefont
  {Caspani}, \citenamefont {Shalaby}, \citenamefont {Giguere}, \citenamefont
  {Lotti}, \citenamefont {Couairon}, \citenamefont {L{\'e}gar{\'e}},
  \citenamefont {Ozaki} \emph {et~al.}}]{clerici2013wavelength}%
  \BibitemOpen
  \bibfield  {author} {\bibinfo {author} {\bibfnamefont {M.}~\bibnamefont
  {Clerici}}, \bibinfo {author} {\bibfnamefont {M.}~\bibnamefont {Peccianti}},
  \bibinfo {author} {\bibfnamefont {B.~E.}\ \bibnamefont {Schmidt}}, \bibinfo
  {author} {\bibfnamefont {L.}~\bibnamefont {Caspani}}, \bibinfo {author}
  {\bibfnamefont {M.}~\bibnamefont {Shalaby}}, \bibinfo {author} {\bibfnamefont
  {M.}~\bibnamefont {Giguere}}, \bibinfo {author} {\bibfnamefont
  {A.}~\bibnamefont {Lotti}}, \bibinfo {author} {\bibfnamefont
  {A.}~\bibnamefont {Couairon}}, \bibinfo {author} {\bibfnamefont
  {F.}~\bibnamefont {L{\'e}gar{\'e}}}, \bibinfo {author} {\bibfnamefont
  {T.}~\bibnamefont {Ozaki}}, \emph {et~al.},\ }\bibfield  {title} {\bibinfo
  {title} {Wavelength scaling of terahertz generation by gas ionization},\
  }\href@noop {} {\bibfield  {journal} {\bibinfo  {journal} {Physical Review
  Letters}\ }\textbf {\bibinfo {volume} {110}},\ \bibinfo {pages} {253901}
  (\bibinfo {year} {2013})}\BibitemShut {NoStop}%
\bibitem [{\citenamefont {de~Alaiza~Mart{\'\i}nez}\ \emph
  {et~al.}(2015)\citenamefont {de~Alaiza~Mart{\'\i}nez}, \citenamefont
  {Babushkin}, \citenamefont {Berg{\'e}}, \citenamefont {Skupin}, \citenamefont
  {Cabrera-Granado}, \citenamefont {K{\"o}hler}, \citenamefont {Morgner},
  \citenamefont {Husakou},\ and\ \citenamefont {Herrmann}}]{de2015boosting}%
  \BibitemOpen
  \bibfield  {author} {\bibinfo {author} {\bibfnamefont {P.~G.}\ \bibnamefont
  {de~Alaiza~Mart{\'\i}nez}}, \bibinfo {author} {\bibfnamefont
  {I.}~\bibnamefont {Babushkin}}, \bibinfo {author} {\bibfnamefont
  {L.}~\bibnamefont {Berg{\'e}}}, \bibinfo {author} {\bibfnamefont
  {S.}~\bibnamefont {Skupin}}, \bibinfo {author} {\bibfnamefont
  {E.}~\bibnamefont {Cabrera-Granado}}, \bibinfo {author} {\bibfnamefont
  {C.}~\bibnamefont {K{\"o}hler}}, \bibinfo {author} {\bibfnamefont
  {U.}~\bibnamefont {Morgner}}, \bibinfo {author} {\bibfnamefont
  {A.}~\bibnamefont {Husakou}},\ and\ \bibinfo {author} {\bibfnamefont
  {J.}~\bibnamefont {Herrmann}},\ }\bibfield  {title} {\bibinfo {title}
  {Boosting terahertz generation in laser-field ionized gases using a sawtooth
  wave shape},\ }\href@noop {} {\bibfield  {journal} {\bibinfo  {journal}
  {Physical Review Letters}\ }\textbf {\bibinfo {volume} {114}},\ \bibinfo
  {pages} {183901} (\bibinfo {year} {2015})}\BibitemShut {NoStop}%
\bibitem [{\citenamefont {Zhang}\ \emph {et~al.}(2016)\citenamefont {Zhang},
  \citenamefont {Chen}, \citenamefont {Chen}, \citenamefont {Zhang},
  \citenamefont {Yu}, \citenamefont {Sheng},\ and\ \citenamefont
  {Zhang}}]{zhang2016controllable}%
  \BibitemOpen
  \bibfield  {author} {\bibinfo {author} {\bibfnamefont {Z.}~\bibnamefont
  {Zhang}}, \bibinfo {author} {\bibfnamefont {Y.}~\bibnamefont {Chen}},
  \bibinfo {author} {\bibfnamefont {M.}~\bibnamefont {Chen}}, \bibinfo {author}
  {\bibfnamefont {Z.}~\bibnamefont {Zhang}}, \bibinfo {author} {\bibfnamefont
  {J.}~\bibnamefont {Yu}}, \bibinfo {author} {\bibfnamefont {Z.}~\bibnamefont
  {Sheng}},\ and\ \bibinfo {author} {\bibfnamefont {J.}~\bibnamefont {Zhang}},\
  }\bibfield  {title} {\bibinfo {title} {Controllable terahertz radiation from
  a linear-dipole array formed by a two-color laser filament in air},\
  }\href@noop {} {\bibfield  {journal} {\bibinfo  {journal} {Physical Review
  Letters}\ }\textbf {\bibinfo {volume} {117}},\ \bibinfo {pages} {243901}
  (\bibinfo {year} {2016})}\BibitemShut {NoStop}%
\bibitem [{\citenamefont {Yoo}\ \emph {et~al.}(2016)\citenamefont {Yoo},
  \citenamefont {Kuk}, \citenamefont {Zhong},\ and\ \citenamefont
  {Kim}}]{yoo2016generation}%
  \BibitemOpen
  \bibfield  {author} {\bibinfo {author} {\bibfnamefont {Y.-J.}\ \bibnamefont
  {Yoo}}, \bibinfo {author} {\bibfnamefont {D.}~\bibnamefont {Kuk}}, \bibinfo
  {author} {\bibfnamefont {Z.}~\bibnamefont {Zhong}},\ and\ \bibinfo {author}
  {\bibfnamefont {K.~Y.}\ \bibnamefont {Kim}},\ }\bibfield  {title} {\bibinfo
  {title} {Generation and characterization of strong terahertz fields from khz
  laser filamentation},\ }\href@noop {} {\bibfield  {journal} {\bibinfo
  {journal} {IEEE Journal of Selected Topics in Quantum Electronics}\ }\textbf
  {\bibinfo {volume} {23}},\ \bibinfo {pages} {1} (\bibinfo {year}
  {2016})}\BibitemShut {NoStop}%
\bibitem [{\citenamefont {Andreeva}\ \emph {et~al.}(2016)\citenamefont
  {Andreeva}, \citenamefont {Kosareva}, \citenamefont {Panov}, \citenamefont
  {Shipilo}, \citenamefont {Solyankin}, \citenamefont {Esaulkov}, \citenamefont
  {de~Alaiza~Mart{\'\i}nez}, \citenamefont {Shkurinov}, \citenamefont
  {Makarov}, \citenamefont {Berg{\'e}} \emph
  {et~al.}}]{andreeva2016ultrabroad}%
  \BibitemOpen
  \bibfield  {author} {\bibinfo {author} {\bibfnamefont {V.~A.}\ \bibnamefont
  {Andreeva}}, \bibinfo {author} {\bibfnamefont {O.~G.}\ \bibnamefont
  {Kosareva}}, \bibinfo {author} {\bibfnamefont {N.~A.}\ \bibnamefont {Panov}},
  \bibinfo {author} {\bibfnamefont {D.~E.}\ \bibnamefont {Shipilo}}, \bibinfo
  {author} {\bibfnamefont {P.~M.}\ \bibnamefont {Solyankin}}, \bibinfo {author}
  {\bibfnamefont {M.~N.}\ \bibnamefont {Esaulkov}}, \bibinfo {author}
  {\bibfnamefont {P.~G.}\ \bibnamefont {de~Alaiza~Mart{\'\i}nez}}, \bibinfo
  {author} {\bibfnamefont {A.~P.}\ \bibnamefont {Shkurinov}}, \bibinfo {author}
  {\bibfnamefont {V.~A.}\ \bibnamefont {Makarov}}, \bibinfo {author}
  {\bibfnamefont {L.}~\bibnamefont {Berg{\'e}}}, \emph {et~al.},\ }\bibfield
  {title} {\bibinfo {title} {Ultrabroad terahertz spectrum generation from an
  air-based filament plasma},\ }\href@noop {} {\bibfield  {journal} {\bibinfo
  {journal} {Physical Review Letters}\ }\textbf {\bibinfo {volume} {116}},\
  \bibinfo {pages} {063902} (\bibinfo {year} {2016})}\BibitemShut {NoStop}%
\bibitem [{\citenamefont {Zhang}\ \emph {et~al.}(2018)\citenamefont {Zhang},
  \citenamefont {Chen}, \citenamefont {Cui}, \citenamefont {He}, \citenamefont
  {Chen}, \citenamefont {Zhang}, \citenamefont {Yu}, \citenamefont {Chen},
  \citenamefont {Sheng},\ and\ \citenamefont {Zhang}}]{zhang2018manipulation}%
  \BibitemOpen
  \bibfield  {author} {\bibinfo {author} {\bibfnamefont {Z.}~\bibnamefont
  {Zhang}}, \bibinfo {author} {\bibfnamefont {Y.}~\bibnamefont {Chen}},
  \bibinfo {author} {\bibfnamefont {S.}~\bibnamefont {Cui}}, \bibinfo {author}
  {\bibfnamefont {F.}~\bibnamefont {He}}, \bibinfo {author} {\bibfnamefont
  {M.}~\bibnamefont {Chen}}, \bibinfo {author} {\bibfnamefont {Z.}~\bibnamefont
  {Zhang}}, \bibinfo {author} {\bibfnamefont {J.}~\bibnamefont {Yu}}, \bibinfo
  {author} {\bibfnamefont {L.}~\bibnamefont {Chen}}, \bibinfo {author}
  {\bibfnamefont {Z.}~\bibnamefont {Sheng}},\ and\ \bibinfo {author}
  {\bibfnamefont {J.}~\bibnamefont {Zhang}},\ }\bibfield  {title} {\bibinfo
  {title} {Manipulation of polarizations for broadband terahertz waves emitted
  from laser plasma filaments},\ }\href@noop {} {\bibfield  {journal} {\bibinfo
   {journal} {Nature Photonics}\ }\textbf {\bibinfo {volume} {12}},\ \bibinfo
  {pages} {554} (\bibinfo {year} {2018})}\BibitemShut {NoStop}%
\bibitem [{\citenamefont {Yoo}\ \emph {et~al.}(2019)\citenamefont {Yoo},
  \citenamefont {Jang},\ and\ \citenamefont {Kim}}]{yoo2019highly}%
  \BibitemOpen
  \bibfield  {author} {\bibinfo {author} {\bibfnamefont {Y.-J.}\ \bibnamefont
  {Yoo}}, \bibinfo {author} {\bibfnamefont {D.}~\bibnamefont {Jang}},\ and\
  \bibinfo {author} {\bibfnamefont {K.~Y.}\ \bibnamefont {Kim}},\ }\bibfield
  {title} {\bibinfo {title} {Highly enhanced terahertz conversion by two-color
  laser filamentation at low gas pressures},\ }\href@noop {} {\bibfield
  {journal} {\bibinfo  {journal} {Optics Express}\ }\textbf {\bibinfo {volume}
  {27}},\ \bibinfo {pages} {22663} (\bibinfo {year} {2019})}\BibitemShut
  {NoStop}%
\bibitem [{\citenamefont {Koulouklidis}\ \emph {et~al.}(2020)\citenamefont
  {Koulouklidis}, \citenamefont {Gollner}, \citenamefont {Shumakova},
  \citenamefont {Fedorov}, \citenamefont {Pug{\v{z}}lys}, \citenamefont
  {Baltu{\v{s}}ka},\ and\ \citenamefont
  {Tzortzakis}}]{koulouklidis2020observation}%
  \BibitemOpen
  \bibfield  {author} {\bibinfo {author} {\bibfnamefont {A.~D.}\ \bibnamefont
  {Koulouklidis}}, \bibinfo {author} {\bibfnamefont {C.}~\bibnamefont
  {Gollner}}, \bibinfo {author} {\bibfnamefont {V.}~\bibnamefont {Shumakova}},
  \bibinfo {author} {\bibfnamefont {V.~Y.}\ \bibnamefont {Fedorov}}, \bibinfo
  {author} {\bibfnamefont {A.}~\bibnamefont {Pug{\v{z}}lys}}, \bibinfo {author}
  {\bibfnamefont {A.}~\bibnamefont {Baltu{\v{s}}ka}},\ and\ \bibinfo {author}
  {\bibfnamefont {S.}~\bibnamefont {Tzortzakis}},\ }\bibfield  {title}
  {\bibinfo {title} {Observation of extremely efficient terahertz generation
  from mid-infrared two-color laser filaments},\ }\href@noop {} {\bibfield
  {journal} {\bibinfo  {journal} {Nature Communications}\ }\textbf {\bibinfo
  {volume} {11}},\ \bibinfo {pages} {292} (\bibinfo {year} {2020})}\BibitemShut
  {NoStop}%
\bibitem [{\citenamefont {Buldt}\ \emph {et~al.}(2021)\citenamefont {Buldt},
  \citenamefont {Stark}, \citenamefont {M{\"u}ller}, \citenamefont {Grebing},
  \citenamefont {Jauregui},\ and\ \citenamefont {Limpert}}]{buldt2021gas}%
  \BibitemOpen
  \bibfield  {author} {\bibinfo {author} {\bibfnamefont {J.}~\bibnamefont
  {Buldt}}, \bibinfo {author} {\bibfnamefont {H.}~\bibnamefont {Stark}},
  \bibinfo {author} {\bibfnamefont {M.}~\bibnamefont {M{\"u}ller}}, \bibinfo
  {author} {\bibfnamefont {C.}~\bibnamefont {Grebing}}, \bibinfo {author}
  {\bibfnamefont {C.}~\bibnamefont {Jauregui}},\ and\ \bibinfo {author}
  {\bibfnamefont {J.}~\bibnamefont {Limpert}},\ }\bibfield  {title} {\bibinfo
  {title} {Gas-plasma-based generation of broadband terahertz radiation with
  640 mw average power},\ }\href@noop {} {\bibfield  {journal} {\bibinfo
  {journal} {Optics Letters}\ }\textbf {\bibinfo {volume} {46}},\ \bibinfo
  {pages} {5256} (\bibinfo {year} {2021})}\BibitemShut {NoStop}%
\bibitem [{\citenamefont {Kim}\ \emph {et~al.}(2008)\citenamefont {Kim},
  \citenamefont {Taylor}, \citenamefont {Glownia},\ and\ \citenamefont
  {Rodriguez}}]{kim2008coherent}%
  \BibitemOpen
  \bibfield  {author} {\bibinfo {author} {\bibfnamefont {K.~Y.}\ \bibnamefont
  {Kim}}, \bibinfo {author} {\bibfnamefont {A.~J.}\ \bibnamefont {Taylor}},
  \bibinfo {author} {\bibfnamefont {J.~H.}\ \bibnamefont {Glownia}},\ and\
  \bibinfo {author} {\bibfnamefont {G.}~\bibnamefont {Rodriguez}},\ }\bibfield
  {title} {\bibinfo {title} {Coherent control of terahertz supercontinuum
  generation in ultrafast laser--gas interactions},\ }\href@noop {} {\bibfield
  {journal} {\bibinfo  {journal} {Nature Photonics}\ }\textbf {\bibinfo
  {volume} {2}},\ \bibinfo {pages} {605} (\bibinfo {year} {2008})}\BibitemShut
  {NoStop}%
\bibitem [{\citenamefont {Johnson}\ \emph {et~al.}(2013)\citenamefont
  {Johnson}, \citenamefont {Palastro}, \citenamefont {Antonsen},\ and\
  \citenamefont {Kim}}]{johnson2013thz}%
  \BibitemOpen
  \bibfield  {author} {\bibinfo {author} {\bibfnamefont {L.~A.}\ \bibnamefont
  {Johnson}}, \bibinfo {author} {\bibfnamefont {J.~P.}\ \bibnamefont
  {Palastro}}, \bibinfo {author} {\bibfnamefont {T.~M.}\ \bibnamefont
  {Antonsen}},\ and\ \bibinfo {author} {\bibfnamefont {K.~Y.}\ \bibnamefont
  {Kim}},\ }\bibfield  {title} {\bibinfo {title} {Thz generation by optical
  cherenkov emission from ionizing two-color laser pulses},\ }\href@noop {}
  {\bibfield  {journal} {\bibinfo  {journal} {Physical Review A}\ }\textbf
  {\bibinfo {volume} {88}},\ \bibinfo {pages} {063804} (\bibinfo {year}
  {2013})}\BibitemShut {NoStop}%
\bibitem [{\citenamefont {Palastro}\ \emph {et~al.}(2018)\citenamefont
  {Palastro}, \citenamefont {Turnbull}, \citenamefont {Bahk}, \citenamefont
  {Follett}, \citenamefont {Shaw}, \citenamefont {Haberberger}, \citenamefont
  {Bromage},\ and\ \citenamefont {Froula}}]{palastro2018ionization}%
  \BibitemOpen
  \bibfield  {author} {\bibinfo {author} {\bibfnamefont {J.~P.}\ \bibnamefont
  {Palastro}}, \bibinfo {author} {\bibfnamefont {D.}~\bibnamefont {Turnbull}},
  \bibinfo {author} {\bibfnamefont {S.~W.}\ \bibnamefont {Bahk}}, \bibinfo
  {author} {\bibfnamefont {R.~K.}\ \bibnamefont {Follett}}, \bibinfo {author}
  {\bibfnamefont {J.~L.}\ \bibnamefont {Shaw}}, \bibinfo {author}
  {\bibfnamefont {D.}~\bibnamefont {Haberberger}}, \bibinfo {author}
  {\bibfnamefont {J.}~\bibnamefont {Bromage}},\ and\ \bibinfo {author}
  {\bibfnamefont {D.~H.}\ \bibnamefont {Froula}},\ }\bibfield  {title}
  {\bibinfo {title} {Ionization waves of arbitrary velocity driven by a flying
  focus},\ }\href@noop {} {\bibfield  {journal} {\bibinfo  {journal} {Physical
  Review A}\ }\textbf {\bibinfo {volume} {97}},\ \bibinfo {pages} {033835}
  (\bibinfo {year} {2018})}\BibitemShut {NoStop}%
\bibitem [{\citenamefont {Palastro}\ \emph {et~al.}(2020)\citenamefont
  {Palastro}, \citenamefont {Shaw}, \citenamefont {Franke}, \citenamefont
  {Ramsey}, \citenamefont {Simpson},\ and\ \citenamefont
  {Froula}}]{palastro2020dephasingless}%
  \BibitemOpen
  \bibfield  {author} {\bibinfo {author} {\bibfnamefont {J.~P.}\ \bibnamefont
  {Palastro}}, \bibinfo {author} {\bibfnamefont {J.~L.}\ \bibnamefont {Shaw}},
  \bibinfo {author} {\bibfnamefont {P.}~\bibnamefont {Franke}}, \bibinfo
  {author} {\bibfnamefont {D.}~\bibnamefont {Ramsey}}, \bibinfo {author}
  {\bibfnamefont {T.~T.}\ \bibnamefont {Simpson}},\ and\ \bibinfo {author}
  {\bibfnamefont {D.~H.}\ \bibnamefont {Froula}},\ }\bibfield  {title}
  {\bibinfo {title} {Dephasingless laser wakefield acceleration},\ }\href@noop
  {} {\bibfield  {journal} {\bibinfo  {journal} {Physical Review Letters}\
  }\textbf {\bibinfo {volume} {124}},\ \bibinfo {pages} {134802} (\bibinfo
  {year} {2020})}\BibitemShut {NoStop}%
\bibitem [{\citenamefont {Simpson}\ \emph {et~al.}(2020)\citenamefont
  {Simpson}, \citenamefont {Ramsey}, \citenamefont {Franke}, \citenamefont
  {Vafaei-Najafabadi}, \citenamefont {Turnbull}, \citenamefont {Froula},\ and\
  \citenamefont {Palastro}}]{simpson2020nonlinear}%
  \BibitemOpen
  \bibfield  {author} {\bibinfo {author} {\bibfnamefont {T.~T.}\ \bibnamefont
  {Simpson}}, \bibinfo {author} {\bibfnamefont {D.}~\bibnamefont {Ramsey}},
  \bibinfo {author} {\bibfnamefont {P.}~\bibnamefont {Franke}}, \bibinfo
  {author} {\bibfnamefont {N.}~\bibnamefont {Vafaei-Najafabadi}}, \bibinfo
  {author} {\bibfnamefont {D.}~\bibnamefont {Turnbull}}, \bibinfo {author}
  {\bibfnamefont {D.~H.}\ \bibnamefont {Froula}},\ and\ \bibinfo {author}
  {\bibfnamefont {J.~P.}\ \bibnamefont {Palastro}},\ }\bibfield  {title}
  {\bibinfo {title} {Nonlinear spatiotemporal control of laser intensity},\
  }\href@noop {} {\bibfield  {journal} {\bibinfo  {journal} {Optics Express}\
  }\textbf {\bibinfo {volume} {28}},\ \bibinfo {pages} {38516} (\bibinfo {year}
  {2020})}\BibitemShut {NoStop}%
\bibitem [{\citenamefont {Franke}\ \emph {et~al.}(2021)\citenamefont {Franke},
  \citenamefont {Ramsey}, \citenamefont {Simpson}, \citenamefont {Turnbull},
  \citenamefont {Froula},\ and\ \citenamefont {Palastro}}]{franke2021optical}%
  \BibitemOpen
  \bibfield  {author} {\bibinfo {author} {\bibfnamefont {P.}~\bibnamefont
  {Franke}}, \bibinfo {author} {\bibfnamefont {D.}~\bibnamefont {Ramsey}},
  \bibinfo {author} {\bibfnamefont {T.~T.}\ \bibnamefont {Simpson}}, \bibinfo
  {author} {\bibfnamefont {D.}~\bibnamefont {Turnbull}}, \bibinfo {author}
  {\bibfnamefont {D.~H.}\ \bibnamefont {Froula}},\ and\ \bibinfo {author}
  {\bibfnamefont {J.~P.}\ \bibnamefont {Palastro}},\ }\bibfield  {title}
  {\bibinfo {title} {Optical shock-enhanced self-photon acceleration},\
  }\href@noop {} {\bibfield  {journal} {\bibinfo  {journal} {Physical Review
  A}\ }\textbf {\bibinfo {volume} {104}},\ \bibinfo {pages} {043520} (\bibinfo
  {year} {2021})}\BibitemShut {NoStop}%
\bibitem [{\citenamefont {Froula}\ \emph {et~al.}(2018)\citenamefont {Froula},
  \citenamefont {Turnbull}, \citenamefont {Davies}, \citenamefont {Kessler},
  \citenamefont {Haberberger}, \citenamefont {Palastro}, \citenamefont {Bahk},
  \citenamefont {Begishev}, \citenamefont {Boni}, \citenamefont {Bucht} \emph
  {et~al.}}]{froula2018spatiotemporal}%
  \BibitemOpen
  \bibfield  {author} {\bibinfo {author} {\bibfnamefont {D.~H.}\ \bibnamefont
  {Froula}}, \bibinfo {author} {\bibfnamefont {D.}~\bibnamefont {Turnbull}},
  \bibinfo {author} {\bibfnamefont {A.~S.}\ \bibnamefont {Davies}}, \bibinfo
  {author} {\bibfnamefont {T.~J.}\ \bibnamefont {Kessler}}, \bibinfo {author}
  {\bibfnamefont {D.}~\bibnamefont {Haberberger}}, \bibinfo {author}
  {\bibfnamefont {J.~P.}\ \bibnamefont {Palastro}}, \bibinfo {author}
  {\bibfnamefont {S.-W.}\ \bibnamefont {Bahk}}, \bibinfo {author}
  {\bibfnamefont {I.~A.}\ \bibnamefont {Begishev}}, \bibinfo {author}
  {\bibfnamefont {R.}~\bibnamefont {Boni}}, \bibinfo {author} {\bibfnamefont
  {S.}~\bibnamefont {Bucht}}, \emph {et~al.},\ }\bibfield  {title} {\bibinfo
  {title} {Spatiotemporal control of laser intensity},\ }\href@noop {}
  {\bibfield  {journal} {\bibinfo  {journal} {Nature Photonics}\ }\textbf
  {\bibinfo {volume} {12}},\ \bibinfo {pages} {262} (\bibinfo {year}
  {2018})}\BibitemShut {NoStop}%
\bibitem [{\citenamefont {Sainte-Marie}\ \emph {et~al.}(2017)\citenamefont
  {Sainte-Marie}, \citenamefont {Gobert},\ and\ \citenamefont
  {Quere}}]{sainte2017controlling}%
  \BibitemOpen
  \bibfield  {author} {\bibinfo {author} {\bibfnamefont {A.}~\bibnamefont
  {Sainte-Marie}}, \bibinfo {author} {\bibfnamefont {O.}~\bibnamefont
  {Gobert}},\ and\ \bibinfo {author} {\bibfnamefont {F.}~\bibnamefont
  {Quere}},\ }\bibfield  {title} {\bibinfo {title} {Controlling the velocity of
  ultrashort light pulses in vacuum through spatio-temporal couplings},\
  }\href@noop {} {\bibfield  {journal} {\bibinfo  {journal} {Optica}\ }\textbf
  {\bibinfo {volume} {4}},\ \bibinfo {pages} {1298} (\bibinfo {year}
  {2017})}\BibitemShut {NoStop}%
\bibitem [{\citenamefont {Simpson}\ \emph {et~al.}(2022)\citenamefont
  {Simpson}, \citenamefont {Ramsey}, \citenamefont {Franke}, \citenamefont
  {Weichman}, \citenamefont {Ambat}, \citenamefont {Turnbull}, \citenamefont
  {Froula},\ and\ \citenamefont {Palastro}}]{simpson2022spatiotemporal}%
  \BibitemOpen
  \bibfield  {author} {\bibinfo {author} {\bibfnamefont {T.~T.}\ \bibnamefont
  {Simpson}}, \bibinfo {author} {\bibfnamefont {D.}~\bibnamefont {Ramsey}},
  \bibinfo {author} {\bibfnamefont {P.}~\bibnamefont {Franke}}, \bibinfo
  {author} {\bibfnamefont {K.}~\bibnamefont {Weichman}}, \bibinfo {author}
  {\bibfnamefont {M.~V.}\ \bibnamefont {Ambat}}, \bibinfo {author}
  {\bibfnamefont {D.}~\bibnamefont {Turnbull}}, \bibinfo {author}
  {\bibfnamefont {D.~H.}\ \bibnamefont {Froula}},\ and\ \bibinfo {author}
  {\bibfnamefont {J.~P.}\ \bibnamefont {Palastro}},\ }\bibfield  {title}
  {\bibinfo {title} {Spatiotemporal control of laser intensity through
  cross-phase modulation},\ }\href@noop {} {\bibfield  {journal} {\bibinfo
  {journal} {Optics Express}\ }\textbf {\bibinfo {volume} {30}},\ \bibinfo
  {pages} {9878} (\bibinfo {year} {2022})}\BibitemShut {NoStop}%
\bibitem [{\citenamefont {Turnbull}\ \emph
  {et~al.}(2018{\natexlab{a}})\citenamefont {Turnbull}, \citenamefont {Franke},
  \citenamefont {Katz}, \citenamefont {Palastro}, \citenamefont {Begishev},
  \citenamefont {Boni}, \citenamefont {Bromage}, \citenamefont {Milder},
  \citenamefont {Shaw},\ and\ \citenamefont {Froula}}]{turnbull2018ionization}%
  \BibitemOpen
  \bibfield  {author} {\bibinfo {author} {\bibfnamefont {D.}~\bibnamefont
  {Turnbull}}, \bibinfo {author} {\bibfnamefont {P.}~\bibnamefont {Franke}},
  \bibinfo {author} {\bibfnamefont {J.}~\bibnamefont {Katz}}, \bibinfo {author}
  {\bibfnamefont {J.~P.}\ \bibnamefont {Palastro}}, \bibinfo {author}
  {\bibfnamefont {I.~A.}\ \bibnamefont {Begishev}}, \bibinfo {author}
  {\bibfnamefont {R.}~\bibnamefont {Boni}}, \bibinfo {author} {\bibfnamefont
  {J.}~\bibnamefont {Bromage}}, \bibinfo {author} {\bibfnamefont {A.~L.}\
  \bibnamefont {Milder}}, \bibinfo {author} {\bibfnamefont {J.~L.}\
  \bibnamefont {Shaw}},\ and\ \bibinfo {author} {\bibfnamefont {D.~H.}\
  \bibnamefont {Froula}},\ }\bibfield  {title} {\bibinfo {title} {Ionization
  waves of arbitrary velocity},\ }\href@noop {} {\bibfield  {journal} {\bibinfo
   {journal} {Physical Review Letters}\ }\textbf {\bibinfo {volume} {120}},\
  \bibinfo {pages} {225001} (\bibinfo {year} {2018}{\natexlab{a}})}\BibitemShut
  {NoStop}%
\bibitem [{\citenamefont {Franke}\ \emph {et~al.}(2019)\citenamefont {Franke},
  \citenamefont {Turnbull}, \citenamefont {Katz}, \citenamefont {Palastro},
  \citenamefont {Begishev}, \citenamefont {Bromage}, \citenamefont {Shaw},
  \citenamefont {Boni},\ and\ \citenamefont {Froula}}]{franke2019measurement}%
  \BibitemOpen
  \bibfield  {author} {\bibinfo {author} {\bibfnamefont {P.}~\bibnamefont
  {Franke}}, \bibinfo {author} {\bibfnamefont {D.}~\bibnamefont {Turnbull}},
  \bibinfo {author} {\bibfnamefont {J.}~\bibnamefont {Katz}}, \bibinfo {author}
  {\bibfnamefont {J.~P.}\ \bibnamefont {Palastro}}, \bibinfo {author}
  {\bibfnamefont {I.~A.}\ \bibnamefont {Begishev}}, \bibinfo {author}
  {\bibfnamefont {J.}~\bibnamefont {Bromage}}, \bibinfo {author} {\bibfnamefont
  {J.~L.}\ \bibnamefont {Shaw}}, \bibinfo {author} {\bibfnamefont
  {R.}~\bibnamefont {Boni}},\ and\ \bibinfo {author} {\bibfnamefont {D.~H.}\
  \bibnamefont {Froula}},\ }\bibfield  {title} {\bibinfo {title} {Measurement
  and control of large diameter ionization waves of arbitrary velocity},\
  }\href@noop {} {\bibfield  {journal} {\bibinfo  {journal} {Optics Express}\
  }\textbf {\bibinfo {volume} {27}},\ \bibinfo {pages} {31978} (\bibinfo {year}
  {2019})}\BibitemShut {NoStop}%
\bibitem [{\citenamefont {Howard}\ \emph {et~al.}(2019)\citenamefont {Howard},
  \citenamefont {Turnbull}, \citenamefont {Davies}, \citenamefont {Franke},
  \citenamefont {Froula},\ and\ \citenamefont {Palastro}}]{howard2019photon}%
  \BibitemOpen
  \bibfield  {author} {\bibinfo {author} {\bibfnamefont {A.~J.}\ \bibnamefont
  {Howard}}, \bibinfo {author} {\bibfnamefont {D.}~\bibnamefont {Turnbull}},
  \bibinfo {author} {\bibfnamefont {A.~S.}\ \bibnamefont {Davies}}, \bibinfo
  {author} {\bibfnamefont {P.}~\bibnamefont {Franke}}, \bibinfo {author}
  {\bibfnamefont {D.~H.}\ \bibnamefont {Froula}},\ and\ \bibinfo {author}
  {\bibfnamefont {J.~P.}\ \bibnamefont {Palastro}},\ }\bibfield  {title}
  {\bibinfo {title} {Photon acceleration in a flying focus},\ }\href@noop {}
  {\bibfield  {journal} {\bibinfo  {journal} {Physical Review Letters}\
  }\textbf {\bibinfo {volume} {123}},\ \bibinfo {pages} {124801} (\bibinfo
  {year} {2019})}\BibitemShut {NoStop}%
\bibitem [{\citenamefont {Turnbull}\ \emph
  {et~al.}(2018{\natexlab{b}})\citenamefont {Turnbull}, \citenamefont {Bucht},
  \citenamefont {Davies}, \citenamefont {Haberberger}, \citenamefont {Kessler},
  \citenamefont {Shaw},\ and\ \citenamefont {Froula}}]{turnbull2018raman}%
  \BibitemOpen
  \bibfield  {author} {\bibinfo {author} {\bibfnamefont {D.}~\bibnamefont
  {Turnbull}}, \bibinfo {author} {\bibfnamefont {S.}~\bibnamefont {Bucht}},
  \bibinfo {author} {\bibfnamefont {A.}~\bibnamefont {Davies}}, \bibinfo
  {author} {\bibfnamefont {D.}~\bibnamefont {Haberberger}}, \bibinfo {author}
  {\bibfnamefont {T.}~\bibnamefont {Kessler}}, \bibinfo {author} {\bibfnamefont
  {J.~L.}\ \bibnamefont {Shaw}},\ and\ \bibinfo {author} {\bibfnamefont
  {D.~H.}\ \bibnamefont {Froula}},\ }\bibfield  {title} {\bibinfo {title}
  {Raman amplification with a flying focus},\ }\href@noop {} {\bibfield
  {journal} {\bibinfo  {journal} {Physical Review Letters}\ }\textbf {\bibinfo
  {volume} {120}},\ \bibinfo {pages} {024801} (\bibinfo {year}
  {2018}{\natexlab{b}})}\BibitemShut {NoStop}%
\bibitem [{\citenamefont {Pigeon}\ \emph {et~al.}(2023)\citenamefont {Pigeon},
  \citenamefont {Franke}, \citenamefont {Lim Pac~Chong}, \citenamefont {Katz},
  \citenamefont {Boni}, \citenamefont {Dorrer}, \citenamefont {Palastro},\ and\
  \citenamefont {Froula}}]{pigeon2023}%
  \BibitemOpen
  \bibfield  {author} {\bibinfo {author} {\bibfnamefont {J.~J.}\ \bibnamefont
  {Pigeon}}, \bibinfo {author} {\bibfnamefont {P.}~\bibnamefont {Franke}},
  \bibinfo {author} {\bibfnamefont {M.}~\bibnamefont {Lim Pac~Chong}}, \bibinfo
  {author} {\bibfnamefont {J.}~\bibnamefont {Katz}}, \bibinfo {author}
  {\bibfnamefont {R.}~\bibnamefont {Boni}}, \bibinfo {author} {\bibfnamefont
  {C.}~\bibnamefont {Dorrer}}, \bibinfo {author} {\bibfnamefont {J.~P.}\
  \bibnamefont {Palastro}},\ and\ \bibinfo {author} {\bibfnamefont {D.~H.}\
  \bibnamefont {Froula}},\ }\bibfield  {title} {\bibinfo {title}
  {Interferometric measurements of the focal velocity and effective pulse
  duration of an ultrafast ‘flying focus’}\ }(\bibinfo {organization} {CLEO
  Conference},\ \bibinfo {year} {2023})\BibitemShut {NoStop}%
\bibitem [{\citenamefont {Smartsev}\ \emph {et~al.}(2019)\citenamefont
  {Smartsev}, \citenamefont {Caizergues}, \citenamefont {Oubrerie},
  \citenamefont {Gautier}, \citenamefont {Goddet}, \citenamefont {Tafzi},
  \citenamefont {Phuoc}, \citenamefont {Malka},\ and\ \citenamefont
  {Thaury}}]{smartsev2019axiparabola}%
  \BibitemOpen
  \bibfield  {author} {\bibinfo {author} {\bibfnamefont {S.}~\bibnamefont
  {Smartsev}}, \bibinfo {author} {\bibfnamefont {C.}~\bibnamefont
  {Caizergues}}, \bibinfo {author} {\bibfnamefont {K.}~\bibnamefont
  {Oubrerie}}, \bibinfo {author} {\bibfnamefont {J.}~\bibnamefont {Gautier}},
  \bibinfo {author} {\bibfnamefont {J.-P.}\ \bibnamefont {Goddet}}, \bibinfo
  {author} {\bibfnamefont {A.}~\bibnamefont {Tafzi}}, \bibinfo {author}
  {\bibfnamefont {K.~T.}\ \bibnamefont {Phuoc}}, \bibinfo {author}
  {\bibfnamefont {V.}~\bibnamefont {Malka}},\ and\ \bibinfo {author}
  {\bibfnamefont {C.}~\bibnamefont {Thaury}},\ }\bibfield  {title} {\bibinfo
  {title} {Axiparabola: a long-focal-depth, high-resolution mirror for
  broadband high-intensity lasers},\ }\href@noop {} {\bibfield  {journal}
  {\bibinfo  {journal} {Optics Letters}\ }\textbf {\bibinfo {volume} {44}},\
  \bibinfo {pages} {3414} (\bibinfo {year} {2019})}\BibitemShut {NoStop}%
\bibitem [{\citenamefont {Oubrerie}\ \emph {et~al.}(2022)\citenamefont
  {Oubrerie}, \citenamefont {Andriyash}, \citenamefont {Lahaye}, \citenamefont
  {Smartsev}, \citenamefont {Malka},\ and\ \citenamefont
  {Thaury}}]{oubrerie2022axiparabola}%
  \BibitemOpen
  \bibfield  {author} {\bibinfo {author} {\bibfnamefont {K.}~\bibnamefont
  {Oubrerie}}, \bibinfo {author} {\bibfnamefont {I.~A.}\ \bibnamefont
  {Andriyash}}, \bibinfo {author} {\bibfnamefont {R.}~\bibnamefont {Lahaye}},
  \bibinfo {author} {\bibfnamefont {S.}~\bibnamefont {Smartsev}}, \bibinfo
  {author} {\bibfnamefont {V.}~\bibnamefont {Malka}},\ and\ \bibinfo {author}
  {\bibfnamefont {C.}~\bibnamefont {Thaury}},\ }\bibfield  {title} {\bibinfo
  {title} {Axiparabola: a new tool for high-intensity optics},\ }\href@noop {}
  {\bibfield  {journal} {\bibinfo  {journal} {Journal of Optics}\ }\textbf
  {\bibinfo {volume} {24}},\ \bibinfo {pages} {045503} (\bibinfo {year}
  {2022})}\BibitemShut {NoStop}%
\bibitem [{\citenamefont {Caizergues}\ \emph {et~al.}(2020)\citenamefont
  {Caizergues}, \citenamefont {Smartsev}, \citenamefont {Malka},\ and\
  \citenamefont {Thaury}}]{caizergues2020phase}%
  \BibitemOpen
  \bibfield  {author} {\bibinfo {author} {\bibfnamefont {C.}~\bibnamefont
  {Caizergues}}, \bibinfo {author} {\bibfnamefont {S.}~\bibnamefont
  {Smartsev}}, \bibinfo {author} {\bibfnamefont {V.}~\bibnamefont {Malka}},\
  and\ \bibinfo {author} {\bibfnamefont {C.}~\bibnamefont {Thaury}},\
  }\bibfield  {title} {\bibinfo {title} {Phase-locked laser-wakefield electron
  acceleration},\ }\href@noop {} {\bibfield  {journal} {\bibinfo  {journal}
  {Nature Photonics}\ }\textbf {\bibinfo {volume} {14}},\ \bibinfo {pages}
  {475} (\bibinfo {year} {2020})}\BibitemShut {NoStop}%
\bibitem [{\citenamefont {Ambat}\ \emph {et~al.}(tted)\citenamefont {Ambat},
  \citenamefont {Palastro}, \citenamefont {Simpson}, \citenamefont {Froula},\
  and\ \citenamefont {Shaw}}]{ambat2023}%
  \BibitemOpen
  \bibfield  {author} {\bibinfo {author} {\bibfnamefont {M.~V.}\ \bibnamefont
  {Ambat}}, \bibinfo {author} {\bibfnamefont {J.~P.}\ \bibnamefont {Palastro}},
  \bibinfo {author} {\bibfnamefont {T.~T.}\ \bibnamefont {Simpson}}, \bibinfo
  {author} {\bibfnamefont {D.~H.}\ \bibnamefont {Froula}},\ and\ \bibinfo
  {author} {\bibfnamefont {J.~L.}\ \bibnamefont {Shaw}},\ }\bibfield  {title}
  {\bibinfo {title} {Design of radial group delay for arbitrary-trajectory
  ultrafast flying focus pulses},\ }\href@noop {} {\  (\bibinfo {year} {2023,
  submitted})}\BibitemShut {NoStop}%
\bibitem [{\citenamefont {Oliver}\ \emph {et~al.}(2020)\citenamefont {Oliver},
  \citenamefont {Spaulding}, \citenamefont {Charles}, \citenamefont {Coates},\
  and\ \citenamefont {Froula}}]{oliver2020}%
  \BibitemOpen
  \bibfield  {author} {\bibinfo {author} {\bibfnamefont {J.~B.}\ \bibnamefont
  {Oliver}}, \bibinfo {author} {\bibfnamefont {J.}~\bibnamefont {Spaulding}},
  \bibinfo {author} {\bibfnamefont {B.}~\bibnamefont {Charles}}, \bibinfo
  {author} {\bibfnamefont {D.}~\bibnamefont {Coates}},\ and\ \bibinfo {author}
  {\bibfnamefont {D.~H.}\ \bibnamefont {Froula}},\ }\bibfield  {title}
  {\bibinfo {title} {Deposition of a discontinuous coated surface to form a
  phase-stepped reflected wavefront},\ }\href@noop {} {\bibfield  {journal}
  {\bibinfo  {journal} {LLE Review}\ }\textbf {\bibinfo {volume} {162}}
  (\bibinfo {year} {2020})}\BibitemShut {NoStop}%
\bibitem [{\citenamefont {Kolesik}\ \emph {et~al.}(2002)\citenamefont
  {Kolesik}, \citenamefont {Moloney},\ and\ \citenamefont
  {Mlejnek}}]{kolesik2002unidirectional}%
  \BibitemOpen
  \bibfield  {author} {\bibinfo {author} {\bibfnamefont {M.}~\bibnamefont
  {Kolesik}}, \bibinfo {author} {\bibfnamefont {J.}~\bibnamefont {Moloney}},\
  and\ \bibinfo {author} {\bibfnamefont {M.}~\bibnamefont {Mlejnek}},\
  }\bibfield  {title} {\bibinfo {title} {Unidirectional optical pulse
  propagation equation},\ }\href@noop {} {\bibfield  {journal} {\bibinfo
  {journal} {Physical Review Letters}\ }\textbf {\bibinfo {volume} {89}},\
  \bibinfo {pages} {283902} (\bibinfo {year} {2002})}\BibitemShut {NoStop}%
\bibitem [{\citenamefont {Kolesik}\ and\ \citenamefont
  {Moloney}(2004)}]{kolesik2004nonlinear}%
  \BibitemOpen
  \bibfield  {author} {\bibinfo {author} {\bibfnamefont {M.}~\bibnamefont
  {Kolesik}}\ and\ \bibinfo {author} {\bibfnamefont {J.~V.}\ \bibnamefont
  {Moloney}},\ }\bibfield  {title} {\bibinfo {title} {Nonlinear optical pulse
  propagation simulation: From maxwell’s to unidirectional equations},\
  }\href@noop {} {\bibfield  {journal} {\bibinfo  {journal} {Physical Review
  E}\ }\textbf {\bibinfo {volume} {70}},\ \bibinfo {pages} {036604} (\bibinfo
  {year} {2004})}\BibitemShut {NoStop}%
\bibitem [{\citenamefont {Couairon}\ \emph {et~al.}(2011)\citenamefont
  {Couairon}, \citenamefont {Brambilla}, \citenamefont {Corti}, \citenamefont
  {Majus}, \citenamefont {de~J.~Ram{\'\i}rez-G{\'o}ngora},\ and\ \citenamefont
  {Kolesik}}]{couairon2011practitioner}%
  \BibitemOpen
  \bibfield  {author} {\bibinfo {author} {\bibfnamefont {A.}~\bibnamefont
  {Couairon}}, \bibinfo {author} {\bibfnamefont {E.}~\bibnamefont {Brambilla}},
  \bibinfo {author} {\bibfnamefont {T.}~\bibnamefont {Corti}}, \bibinfo
  {author} {\bibfnamefont {D.}~\bibnamefont {Majus}}, \bibinfo {author}
  {\bibfnamefont {O.}~\bibnamefont {de~J.~Ram{\'\i}rez-G{\'o}ngora}},\ and\
  \bibinfo {author} {\bibfnamefont {M.}~\bibnamefont {Kolesik}},\ }\bibfield
  {title} {\bibinfo {title} {Practitioner’s guide to laser pulse propagation
  models and simulation: Numerical implementation and practical usage of modern
  pulse propagation models},\ }\href@noop {} {\bibfield  {journal} {\bibinfo
  {journal} {The European Physical Journal Special Topics}\ }\textbf {\bibinfo
  {volume} {199}},\ \bibinfo {pages} {5} (\bibinfo {year} {2011})}\BibitemShut
  {NoStop}%
\bibitem [{\citenamefont {Berg{\'e}}\ \emph {et~al.}(2013)\citenamefont
  {Berg{\'e}}, \citenamefont {Skupin}, \citenamefont {K{\"o}hler},
  \citenamefont {Babushkin},\ and\ \citenamefont {Herrmann}}]{berge20133d}%
  \BibitemOpen
  \bibfield  {author} {\bibinfo {author} {\bibfnamefont {L.}~\bibnamefont
  {Berg{\'e}}}, \bibinfo {author} {\bibfnamefont {S.}~\bibnamefont {Skupin}},
  \bibinfo {author} {\bibfnamefont {C.}~\bibnamefont {K{\"o}hler}}, \bibinfo
  {author} {\bibfnamefont {I.}~\bibnamefont {Babushkin}},\ and\ \bibinfo
  {author} {\bibfnamefont {J.}~\bibnamefont {Herrmann}},\ }\bibfield  {title}
  {\bibinfo {title} {3d numerical simulations of thz generation by two-color
  laser filaments},\ }\href@noop {} {\bibfield  {journal} {\bibinfo  {journal}
  {Physical Review Letters}\ }\textbf {\bibinfo {volume} {110}},\ \bibinfo
  {pages} {073901} (\bibinfo {year} {2013})}\BibitemShut {NoStop}%
\bibitem [{\citenamefont {KMLabs}()}]{KMLabs}%
  \BibitemOpen
  \bibfield  {author} {\bibinfo {author} {\bibnamefont {KMLabs}},\ }\href@noop
  {} {\bibinfo {title} {Raea ultrafast amplifier}},\ \bibinfo {howpublished}
  {\url{https://www.kmlabs.com/raea-ultrafast-amplifier}}\BibitemShut {NoStop}%
\bibitem [{\citenamefont {Zahedpour}\ \emph {et~al.}(2015)\citenamefont
  {Zahedpour}, \citenamefont {Wahlstrand},\ and\ \citenamefont
  {Milchberg}}]{zahedpour2015measurement}%
  \BibitemOpen
  \bibfield  {author} {\bibinfo {author} {\bibfnamefont {S.}~\bibnamefont
  {Zahedpour}}, \bibinfo {author} {\bibfnamefont {J.}~\bibnamefont
  {Wahlstrand}},\ and\ \bibinfo {author} {\bibfnamefont {H.}~\bibnamefont
  {Milchberg}},\ }\bibfield  {title} {\bibinfo {title} {Measurement of the
  nonlinear refractive index of air constituents at mid-infrared wavelengths},\
  }\href@noop {} {\bibfield  {journal} {\bibinfo  {journal} {Optics letters}\
  }\textbf {\bibinfo {volume} {40}},\ \bibinfo {pages} {5794} (\bibinfo {year}
  {2015})}\BibitemShut {NoStop}%
\bibitem [{\citenamefont {Sprangle}\ \emph {et~al.}(2004)\citenamefont
  {Sprangle}, \citenamefont {Penano}, \citenamefont {Hafizi},\ and\
  \citenamefont {Kapetanakos}}]{sprangle2004ultrashort}%
  \BibitemOpen
  \bibfield  {author} {\bibinfo {author} {\bibfnamefont {P.}~\bibnamefont
  {Sprangle}}, \bibinfo {author} {\bibfnamefont {J.~R.}\ \bibnamefont
  {Penano}}, \bibinfo {author} {\bibfnamefont {B.}~\bibnamefont {Hafizi}},\
  and\ \bibinfo {author} {\bibfnamefont {C.~A.}\ \bibnamefont {Kapetanakos}},\
  }\bibfield  {title} {\bibinfo {title} {Ultrashort laser pulses and
  electromagnetic pulse generation in air and on dielectric surfaces},\
  }\href@noop {} {\bibfield  {journal} {\bibinfo  {journal} {Physical Review
  E}\ }\textbf {\bibinfo {volume} {69}},\ \bibinfo {pages} {066415} (\bibinfo
  {year} {2004})}\BibitemShut {NoStop}%
\bibitem [{\citenamefont {Peck}\ and\ \citenamefont
  {Fisher}(1964)}]{peck1964dispersion}%
  \BibitemOpen
  \bibfield  {author} {\bibinfo {author} {\bibfnamefont {E.~R.}\ \bibnamefont
  {Peck}}\ and\ \bibinfo {author} {\bibfnamefont {D.~J.}\ \bibnamefont
  {Fisher}},\ }\bibfield  {title} {\bibinfo {title} {Dispersion of argon},\
  }\href@noop {} {\bibfield  {journal} {\bibinfo  {journal} {JOSA}\ }\textbf
  {\bibinfo {volume} {54}},\ \bibinfo {pages} {1362} (\bibinfo {year}
  {1964})}\BibitemShut {NoStop}%
\bibitem [{\citenamefont {You}\ \emph {et~al.}(2012)\citenamefont {You},
  \citenamefont {Oh},\ and\ \citenamefont {Kim}}]{you2012off}%
  \BibitemOpen
  \bibfield  {author} {\bibinfo {author} {\bibfnamefont {Y.~S.}\ \bibnamefont
  {You}}, \bibinfo {author} {\bibfnamefont {T.~I.}\ \bibnamefont {Oh}},\ and\
  \bibinfo {author} {\bibfnamefont {K.~Y.}\ \bibnamefont {Kim}},\ }\bibfield
  {title} {\bibinfo {title} {Off-axis phase-matched terahertz emission from
  two-color laser-induced plasma filaments},\ }\href@noop {} {\bibfield
  {journal} {\bibinfo  {journal} {Physical Review Letters}\ }\textbf {\bibinfo
  {volume} {109}},\ \bibinfo {pages} {183902} (\bibinfo {year}
  {2012})}\BibitemShut {NoStop}%
\bibitem [{\citenamefont {Lee}\ \emph {et~al.}(2023)\citenamefont {Lee},
  \citenamefont {Song}, \citenamefont {Park}, \citenamefont {Kumar},
  \citenamefont {Ersfeld}, \citenamefont {Yoffe}, \citenamefont {Jaroszynski},\
  and\ \citenamefont {Hur}}]{lee2023intense}%
  \BibitemOpen
  \bibfield  {author} {\bibinfo {author} {\bibfnamefont {J.}~\bibnamefont
  {Lee}}, \bibinfo {author} {\bibfnamefont {H.~S.}\ \bibnamefont {Song}},
  \bibinfo {author} {\bibfnamefont {D.}~\bibnamefont {Park}}, \bibinfo {author}
  {\bibfnamefont {M.}~\bibnamefont {Kumar}}, \bibinfo {author} {\bibfnamefont
  {B.}~\bibnamefont {Ersfeld}}, \bibinfo {author} {\bibfnamefont {S.~R.}\
  \bibnamefont {Yoffe}}, \bibinfo {author} {\bibfnamefont {D.~A.}\ \bibnamefont
  {Jaroszynski}},\ and\ \bibinfo {author} {\bibfnamefont {M.~S.}\ \bibnamefont
  {Hur}},\ }\bibfield  {title} {\bibinfo {title} {Intense narrowband terahertz
  pulses produced by obliquely colliding laser pulses in helium gas},\
  }\href@noop {} {\bibfield  {journal} {\bibinfo  {journal} {Physics of
  Plasmas}\ }\textbf {\bibinfo {volume} {30}},\ \bibinfo {pages} {043108}
  (\bibinfo {year} {2023})}\BibitemShut {NoStop}%
\bibitem [{\citenamefont {Yoshii}\ \emph {et~al.}(1997)\citenamefont {Yoshii},
  \citenamefont {Lai}, \citenamefont {Katsouleas}, \citenamefont {Joshi},\ and\
  \citenamefont {Mori}}]{yoshii1997radiation}%
  \BibitemOpen
  \bibfield  {author} {\bibinfo {author} {\bibfnamefont {J.}~\bibnamefont
  {Yoshii}}, \bibinfo {author} {\bibfnamefont {C.~H.}\ \bibnamefont {Lai}},
  \bibinfo {author} {\bibfnamefont {T.}~\bibnamefont {Katsouleas}}, \bibinfo
  {author} {\bibfnamefont {C.}~\bibnamefont {Joshi}},\ and\ \bibinfo {author}
  {\bibfnamefont {W.~B.}\ \bibnamefont {Mori}},\ }\bibfield  {title} {\bibinfo
  {title} {Radiation from cerenkov wakes in a magnetized plasma},\ }\href@noop
  {} {\bibfield  {journal} {\bibinfo  {journal} {Physical Review Letters}\
  }\textbf {\bibinfo {volume} {79}},\ \bibinfo {pages} {4194} (\bibinfo {year}
  {1997})}\BibitemShut {NoStop}%
\bibitem [{\citenamefont {Guizar-Sicairos}\ and\ \citenamefont
  {Guti{\'e}rrez-Vega}(2004)}]{guizar2004computation}%
  \BibitemOpen
  \bibfield  {author} {\bibinfo {author} {\bibfnamefont {M.}~\bibnamefont
  {Guizar-Sicairos}}\ and\ \bibinfo {author} {\bibfnamefont {J.~C.}\
  \bibnamefont {Guti{\'e}rrez-Vega}},\ }\bibfield  {title} {\bibinfo {title}
  {Computation of quasi-discrete hankel transforms of integer order for
  propagating optical wave fields},\ }\href@noop {} {\bibfield  {journal}
  {\bibinfo  {journal} {JOSA A}\ }\textbf {\bibinfo {volume} {21}},\ \bibinfo
  {pages} {53} (\bibinfo {year} {2004})}\BibitemShut {NoStop}%
\bibitem [{\citenamefont {Ammosov}\ \emph {et~al.}(1986)\citenamefont
  {Ammosov}, \citenamefont {Delone},\ and\ \citenamefont
  {Krainov}}]{ammosov1986tunnel}%
  \BibitemOpen
  \bibfield  {author} {\bibinfo {author} {\bibfnamefont {M.~V.}\ \bibnamefont
  {Ammosov}}, \bibinfo {author} {\bibfnamefont {N.~B.}\ \bibnamefont
  {Delone}},\ and\ \bibinfo {author} {\bibfnamefont {V.~P.}\ \bibnamefont
  {Krainov}},\ }\bibfield  {title} {\bibinfo {title} {Tunnel ionization of
  complex atoms and of atomic ions in an alternating electromagnetic field},\
  }\href@noop {} {\bibfield  {journal} {\bibinfo  {journal} {Soviet Journal of
  Experimental and Theoretical Physics}\ }\textbf {\bibinfo {volume} {64}},\
  \bibinfo {pages} {1191} (\bibinfo {year} {1986})}\BibitemShut {NoStop}%
\bibitem [{\citenamefont {Richardson}(2019)}]{richardson20192019}%
  \BibitemOpen
  \bibfield  {author} {\bibinfo {author} {\bibfnamefont {A.~S.}\ \bibnamefont
  {Richardson}},\ }\href@noop {} {\emph {\bibinfo {title} {2019 NRL plasma
  formulary}}},\ \bibinfo {type} {Tech. Rep.}\ (\bibinfo  {institution} {US
  Naval Research Laboratory},\ \bibinfo {year} {2019})\BibitemShut {NoStop}%
\end{thebibliography}%
\end{document}